\documentclass[useAMS,usenatbib]{mn2e}
\usepackage{graphicx}
\usepackage{journal_names}
\usepackage{longtable}

\usepackage{subfig}
\usepackage{amsmath}
\usepackage{amssymb}
\usepackage{pdflscape}

\title[Statistical properties of dwarf novae]{Statistical Properties of Dwarf Novae-type\\Cataclysmic Variables: The Outburst Catalogue}
\author[Deanne L. Coppejans et al.]{Deanne L. Coppejans$^{1}$\thanks{Email: d.coppejans@astro.ru.nl}, Elmar G. K\"{o}rding$^{1}$, Christian Knigge$^{2}$ \newauthor Margaretha L. Pretorius$^{3}$, Patrick A. Woudt$^{4}$, Paul J. Groot$^1$ \newauthor Cameron L. Van Eck$^{1}$, Andrew J. Drake$^5$\\
$^{1}$Department of Astrophysics/IMAPP, Radboud University, P.O. Box 9010, 6500 GL Nijmegen, The Netherlands\\
$^{2}$School of Physics and Astronomy, Southampton University, Highfield, Southampton SO17 1BJ, UK\\
$^{3}$Oxford University, Department of Physics, Denys Wilkinson Building, Keble Road, Oxford, OX1 3RH, UK\\
$^{4}$Astrophysics, Cosmology and Gravity Centre, Department of Astronomy, University of Cape Town,\\Private Bag X3, 7701 Rondebosch, South Africa\\
${^5}$California Institute of Technology, 1200 E. California Blvd, CA 91225, USA}
\begin{document} 

\date{}

\pagerange{\pageref{firstpage}--\pageref{lastpage}} \pubyear{2015}

\maketitle

\label{firstpage}

\begin{abstract}

The Outburst Catalogue contains a wide variety of observational properties for 722 dwarf nova-type (DN) cataclysmic variables (CVs) and 309 CVs of other types from the Catalina Real-time Transient Survey. In particular, it includes the apparent outburst and quiescent $V$-band magnitudes, duty cycles, limits on the recurrence time, upper- and lower-limits on the distance and absolute quiescent magnitudes, colour information, orbital parameters, and X-ray counterparts. These properties were determined by means of a classification script presented in this paper.\\
The DN in the catalogue show a correlation between the outburst duty cycle and the orbital period (and outburst recurrence time), as well as between the quiescent absolute magnitude and the orbital period (and duty cycle).\\
This is the largest sample of dwarf nova properties collected to date. Besides serving as a useful reference for individual systems and a means of selecting objects for targetted studies, it will prove valuable for statistical studies that aim to shed light on the formation and evolution of cataclysmic variables.

\end{abstract}


\section{Introduction}

Cataclysmic Variable stars (CVs) are interacting binary systems which comprise a white dwarf (WD) primary and a red dwarf secondary star (see \citealt{Warner1995} for a review). Mass transfer occurs via Roche-lobe overflow, and accretion onto the surface of the primary can either proceed via an accretion disc (the non-magnetic systems), or via magnetic field line-channeling from the Alf\'{v}en radius if the magnetic field strength is sufficiently high (the magnetic systems).

The main CV subclasses are primarily defined according to their long-term photometric behaviour and the magnetic field strength of the WD. In many of the non-magnetic systems ($B_{\rm WD}\lesssim10^6\,$G), the accretion disc switches between a cool, un-ionised state and a hot, ionised state with a high mass-transfer rate. This is commonly accepted to be caused by a thermal-viscous instability (\citealt{Smak1971, Osaki1974, Hoshi1979}). The interludes where the disc is bright and hot are known as dwarf nova outbursts (hereafter refered to as outbursts). Outbursts typically recur on timescales of days to decades, last for about a week and have outburst amplitudes of 2--8 magnitudes in the optical, but there are large variations between CVs in all three of these properties. The CVs that show these outbursts are known as dwarf novae (DN). Non-magnetic CVs in which the mass-transfer rate from the secondary ($\dot{M}$) is sufficiently high to maintain the disc in a hot state, are the novalikes. Some novalikes are known to show occasional low-states lasting weeks to years (e.g. \citealt{Groot2001,Honeycutt2004}). The polars and intermediate polars form the two classes of magnetic CVs. In intermediate polars ($10^6\lesssim B_{\rm WD}\lesssim10^7\,$G) a partial accretion disc is present (it is truncated at the Alf\'{v}en radius), whereas in polars the magnetic field strength is sufficiently high ($B_{\rm WD}\gtrsim10^7$G) that material is fed directly from the secondary onto magnetic field lines at the point where the magnetic pressure exceeds the ram pressure.

In the last few years a number of discrepancies have emerged in the evolutionary model for (particularly) the non-magnetic CVs. First, it was predicted that there should be a spike in the number of CVs at the minimum orbital period, $P_{\rm orb}^{\rm min}\approx65\,$mins \citep{Paczynski1981,Rappaport1982,Kolb1999,Howell2001}. This period spike has now been detected and confirmed observationally, but at the longer orbital period ($P_{\rm orb}$) of approximately 82 mins \citep{Gaensicke2009, Woudt2012, Drake2014}. A CV initially evolves to shorter $P_{\rm orb}$ via a loss of angular momentum by magnetic braking and gravitational radiation. During this time, the mass-loss rate from the secondary drives it increasingly out of thermal-equilibrium until the thermal-timescale exceeds the mass-loss timescale and it expands in response to mass loss -- thereby increasing $P_{\rm orb}$. Consequently the evolutionary direction changes at $P_{\rm orb}^{\rm min}$, and a large number of CVs is expected at this orbital period (the period spike). In order to reconcile the observed and predicted values for $P_{\rm orb}^{\rm min}$, enhanced angular momentum loss at short orbital periods has been suggested \citep{Knigge2011}. Second, a large population of CVs that have evolved past $P_{\rm orb}^{\rm min}$ (post period-minimum CVs) is predicted. The fraction of post-bounce CVs is expected to be 40-70\%, however, only a few candidates have been identified to date (e.g. \citealt{Littlefair2008}).

Deep, long-term, time-domain surveys offer a solution to these problems, as they are detecting larger, deeper, and less-biased samples of CVs. Examples of these surveys include the Catalina Real-time Transient Survey (CRTS, \citealt{Drake2009}), the Palomar Quest digital synoptic sky survey (PQ, \citealt{Djorgovski2008}) and the Palomar Transient Factory (PTF, \citealt{Law2009,Rau2009}), as well as the upcoming Large Scale Synoptic Telescope (LSST, \citealt{Tyson2002}). The overall strategy of these surveys is to detect variable objects through multi-epoch observations, but the observing cadence, sky coverage and variability criteria differ between the surveys. CVs are detected in large numbers by these surveys due to their range of variability amplitudes and time-scales: from sub-second to minute variability (below $\Delta V\sim10^{-3}$ mag, e.g. \citealt{Woudt2002,Scaringi2014}), orbital modulations on time-scales of minutes to hours ($\Delta V\lesssim3$ mag, e.g. \citealt{Coppejans2014}), to outbursts ($\Delta V\sim8\,$mag, which include dwarf nova outbursts and the longer, brighter superoutbursts -- see \citealt{Warner1995}). Additionally, thermo-nuclear runaways on the surface of the WD (Novae) produce amplitude variations of up to $\Delta V\sim10$ mag (for a review, see \citealt{Bode2010}).

The CRTS, in particular, has detected more than 1000 CVs to date, each with a light curve spanning $\sim8$ to 9 years. We have used these long-term light curves to estimate the duty cycle and constrain the outburst recurrence time for a large sample of CVs. These properties will help constrain the mass-transfer rate and angular momentum loss (AML) rate in the overall population. Better-constrained AML rates will improve evolutionary models and our understanding of the post-bounce population, and help to reconcile the observed and predicted position of the period minimum. 

Our catalogue was constructed with a goal towards answering these, and related, questions that require large samples of CV properties to answer. Additionally it is meant as a reference for individual systems and a means of selecting CVs for targeted campaigns, based on their outburst properties.

In Section \ref{sec:CRTS} we discuss the CRTS survey in detail. Section \ref{sec:code} describes the classification script, and Section \ref{sec:catalogue} outlines the contents of our catalogue. The catalogue is then characterised in Section \ref{sec:discussion}.

\section{CRTS}\label{sec:CRTS}

The Catalina Real-time Transient Survey (CRTS) identifies transients in the data from the Catalina Sky Survey\footnote{http://www.lpl.arizona.edu/css/} (\citealt{Larson1998,Larson2003,Johnson2014css}) -- a photometric survey that searches for Potentially Hazardous Asteroids (PHAs) and Near Earth Objects (NEOs). Three sub-surveys constitute the Catalina Sky Survey, namely the original CSS (Catalina Schmidt Survey), the MLS (Mt. Lemmon Survey) based in Arizona, and the SSS (Siding Spring Survey) in Australia, which ended on 5 July 2014. Each has a dedicated telescope with a 4k back-illuminated, unfiltered CCD camera \citep{Djorgovski2010}. The field of view and typical limiting magnitude for each survey (at $\sim30\,$s integrations) are 8.2\degr and $V\sim19.5$ mag for the CSS, 1.1\degr and $V\sim21.5$ mag for the MLS, and 4\degr and $V\sim19$ mag for the SSS. Together these surveys cover 30,000 deg$^2$ ($-70\degr<\delta<70\degr$, see \citealt{Drake2014} for more details). The Galactic plane ($|b|<15\degr$) is avoided due to overcrowding, as are the Magellanic Clouds.

Each observation consists of four images (frames) that are separated by $\approx$10 minutes. The observing cadence is typically 1-4 times per lunation (depending on the sub-survey and field). Aperture photometry is performed using SExtractor software \citep{Bertin1996} and converted to $V$-band magnitudes using standard stars as described in \citet{Drake2013}.

The CRTS began processing CSS data on 8 November 2007, MLS data on 6 November 2009 and 5 May 2010 for SSS data \citep{Drake2014}. Although there is Catalina Sky Survey data preceeding these dates, the CRTS only use these data in the transient classification process -- any object that was only variable during this time will not have been identified as a CRTS transient.

To identify variability, the CRTS makes catalogues of the objects in each image and compare these to previously recorded magnitudes; they do not use image subtraction. An object needs to pass a number of tests to be classified as a transient. In each set of four frames (one epoch), it needs to be positionally coincident in at least three of the frames. This eliminates high proper motion objects (movement of $>0.1'\,{\rm min}^{-1}$ between frames), which are generally solar system objects. Additionally it needs to be a point source, and cannot be saturated ($V\gtrsim12.5$ mag), or blended (the pixel scale is 2.5$''$).

Objects that pass these tests are then compared to deep co-added image catalogues. There is one co-add per CRTS field and it is the median of 20 images taken at the beginning of the CRTS survey. The co-adds typically reach down to $V\sim21.5$ mag for the CSS, $V\sim22.5-23$ mag for the MLS and $V\sim20$ mag for the SSS.

The criteria for classifying a transient have evolved over the course of the survey to detect more transients and filter out periodic variable stars (see \citealt{Drake2014}). Initially an object needed to be either $\geq2$ mag brighter than the co-add or had to be absent in the co-add \citep{Drake2009}. This requirement changed to $\geq1$ mag brighter than the co-add (or absent in the co-add). Currently, an object needs to be $\geq0.65$ mag brighter than the co-add (or absent in the co-add), with a $\geq3\sigma$ flux change in comparison to its CRTS light curve \citep{Drake2014}. The new criteria have not been applied to previously processed data. The candidate transient is then compared to archival data from the SDSS, the USNO-B (US Naval Oservatory B catalogue, \citealt{Monet2003}) and the Palomar-Quest Synoptic Sky Survey \citep[PQ,][]{Djorgovski2008} in order to discard further artifacts, for example those that were missing in the co-adds because they were blended. As a final check against artifacts, new transients are examined by eye and assigned a classification of CV, supernova (SN), quasar, asteroid or flare star, blazar, AGN, or unknown.  

We now describe the CRTS classification procedure in relation to CVs; for details regarding other classes of transients see \citet{Drake2009,Drake2014}. If available, the classification given in the Virtual Observatory (VO, \citealt{Quinn2004,Borne2013}) is used, otherwise spectra and photometry from SDSS, USNO-B and PQ, along with the CRTS light curves are used in the classification. A number of features are taken into account when classifying an object as a CV. Multiple outbursts, rapid declines, a return to quiescence within a short time, a variable quiescent level, and a blue point-source counterpart in the SDSS, all increase the probability of a transient being classified as a CV. Colour-cuts are not used. Objects that show only one outburst could be either CVs or SN, although CVs are generally brighter on average and more likely to be seen in quiescence. If an object with a single outburst has a background galaxy, it is classified as a SN (as SN are likely to be followed up, misclassifications are generally caught). If it is not clear whether an object with a single outburst is a CV or SN, it is placed in the `CV or SN' category in the CRTS. Note that since the CRTS do not routinely reclassify, a CV that shows another outburst after the classification may still be in the `CV or SN' category. CRTS follow-up photometry and spectroscopy has been performed for some of the CVs \citep{Drake2014}.

The CRTS data is open access, so the images and light curves are available to the public. The discovery date, magnitude, change in magnitude, classification and images from other surveys are also included. This has led to a number of photometric and spectroscopic CV follow-up studies (e.g. \citealt{Thorstensen2012,Woudt2012,Kato2013,Breedt2014,Coppejans2014,Szkody2014} and references therein). These surveys indicate that more than 95\% of objects classified as CVs were correctly classified \citep{Breedt2014,Drake2014}. Any misclassification noted in the ATels or literature is corrected in the CRTS database.

Up to October 2015, the CRTS had detected a total of 10,782 transients. This total includes 1252 CVs, 2570 supernovae, 1476 CVs/supernovae, 638 asteroids/flares, 373 blazars and 2968 AGN. An up-to-date tally is given on the CRTS website\footnote{http://crts.caltech.edu/}.

\section{Classification Script}\label{sec:code}

\begin{figure*}
  \centering
    \includegraphics[width=14cm]{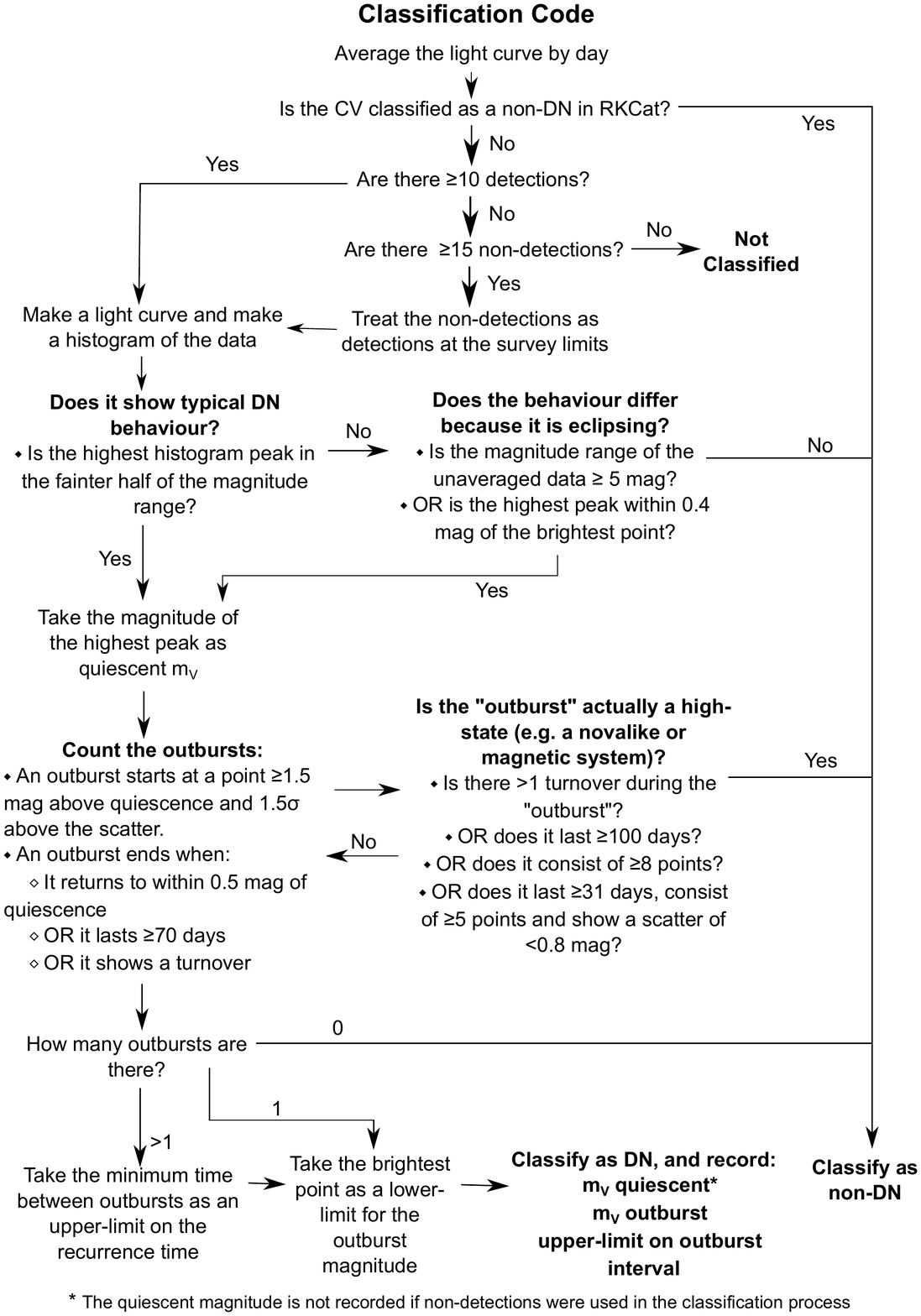}
    \caption{Flowchart depicting the process followed by the classification script, whereby a CV is classified as `DN' or `non-DN' based on the light curve. Outburst properties are subsequently estimated for the DN systems.}
  \label{fig:code_flowchart}
\end{figure*}

The 8--9 year CRTS CV light curves offer a good means to estimate and constrain outburst properties for DN, such as the duty cycle, recurrence time, outburst amplitude, and apparent quiescent and outburst magnitudes.

Due to the difficulties involved with making these estimates from magnitude-limited, irregularly-sampled data, previous estimates for these properties have been made by eye (e.g. \citealt{Breedt2014,Drake2014}). Scripting this procedure is advantageous for a number of reasons. It defines hard classification criteria. This makes it possible to determine completeness, compare the sample to other databases and, importantly, update it when new observations become available. Using a script also makes it possible to trace outbursts and estimate recurrence times.

We have written a script, hereafter refered to as `the classification script', that uses light curves to classify a CV as a `DN' or `non-DN', and subsequently estimate and define limits for the duty cycle, outburst recurrence time, apparent outburst and quiescent magnitudes, and distance. Only those transients that had already been classified as CVs by the CRTS have been run through the script. A flowchart describing the procedure is show in Figure \ref{fig:code_flowchart}.

The CRTS input light curves for the classification script are generated from the published light curves and the observing log. The latter is necessary to determine the times when the objects were not detected (the non-detections/upper-limits are not included in the current CRTS data release). The exact magnitude of the upper-limit is not important for the classification script, so it is recorded at the average limiting magnitude of the survey\footnote{See http://crts.caltech.edu/Telescopes.html} (SSS: $V=19$ mag, MLS: $V=21.5$ mag, CSS: $V=19.5$ mag).

The first steps in the script are to average the light curve, and to determine if there are enough data points to attempt a classification. Each light curve is averaged by day (by set of four CRTS observations separated by 10 minutes) in order to reduce the scatter introduced by eclipses and variability. A minimum of 10 detections, or 15 non-detections, are then required to proceed.

Through a series of tests, the script assigns one of two possible classifications to every CV, namely `DN' or `non-DN'. The former consists of all CVs that show outbursts, while the latter consists of non-outbursting CVs such as polars and novalikes.

\begin{figure}
  \centering
    \includegraphics[width=8cm]{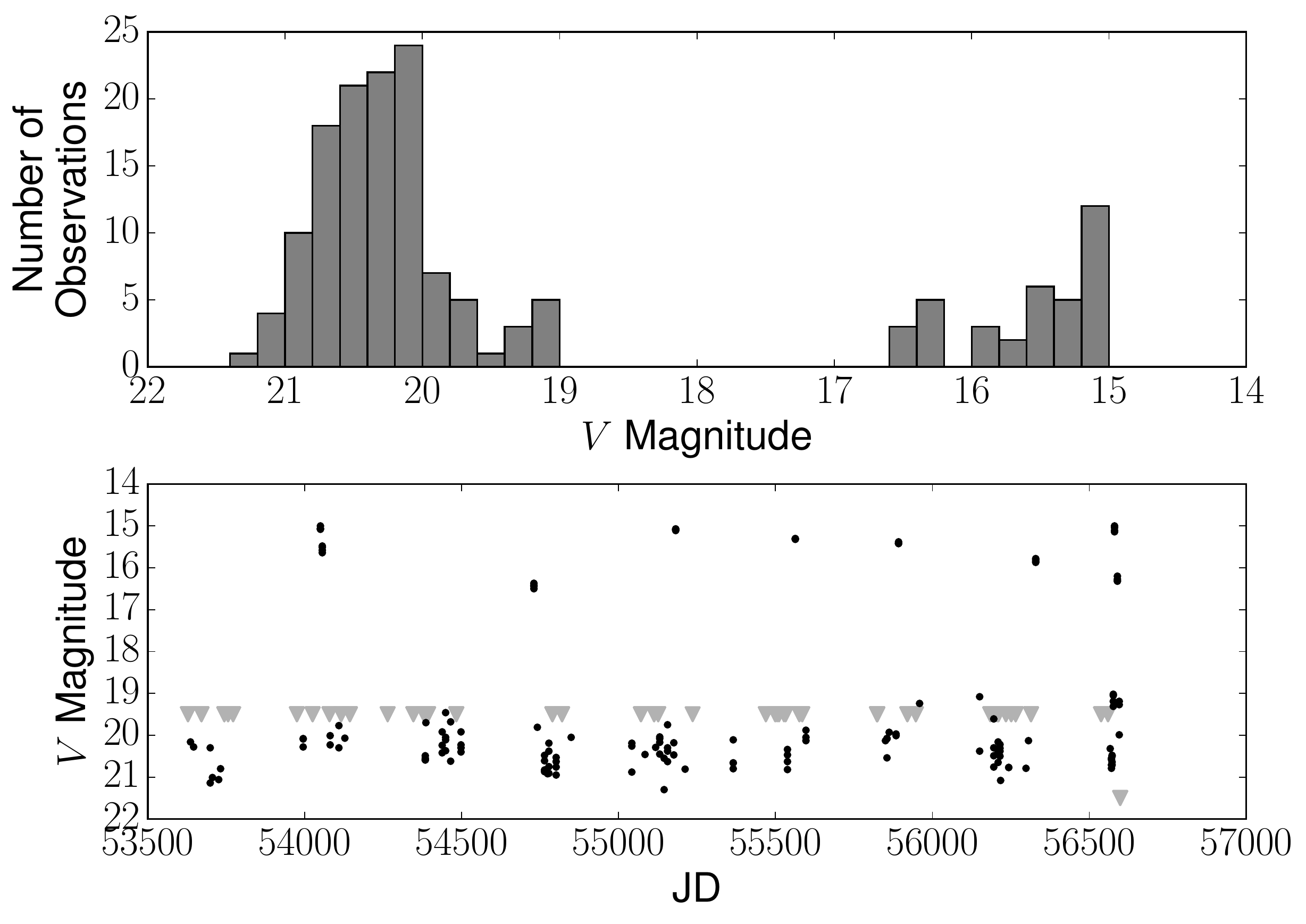}
    \caption{Illustration of the initial classification step (see Figure \ref{fig:code_flowchart}), where the light curve (bottom panel) is histogrammed by magnitude (top panel) in order to determine whether it shows standard DN-type behaviour. In this example the CV was given the initial classification of `DN', because the higher histogram peak is in the fainter half of the magnitude range -- as one would expect from a DN that spends the majority of its time in quiescence. Subsequent tests in the script confirmed the `DN' classification, counted 7 outbursts, and estimated $v_{\rm Q}=$20.1 mag and $v_{\rm O}<15.5$ mag. Note that the non-detections in the lightcurves are shown at the typical detection limit of the survey (currently the CRTS data release does not provide upper limits).}
  \label{fig:lc_standardDN}
\end{figure}

\begin{figure}
  \centering
  \includegraphics[width=8cm]{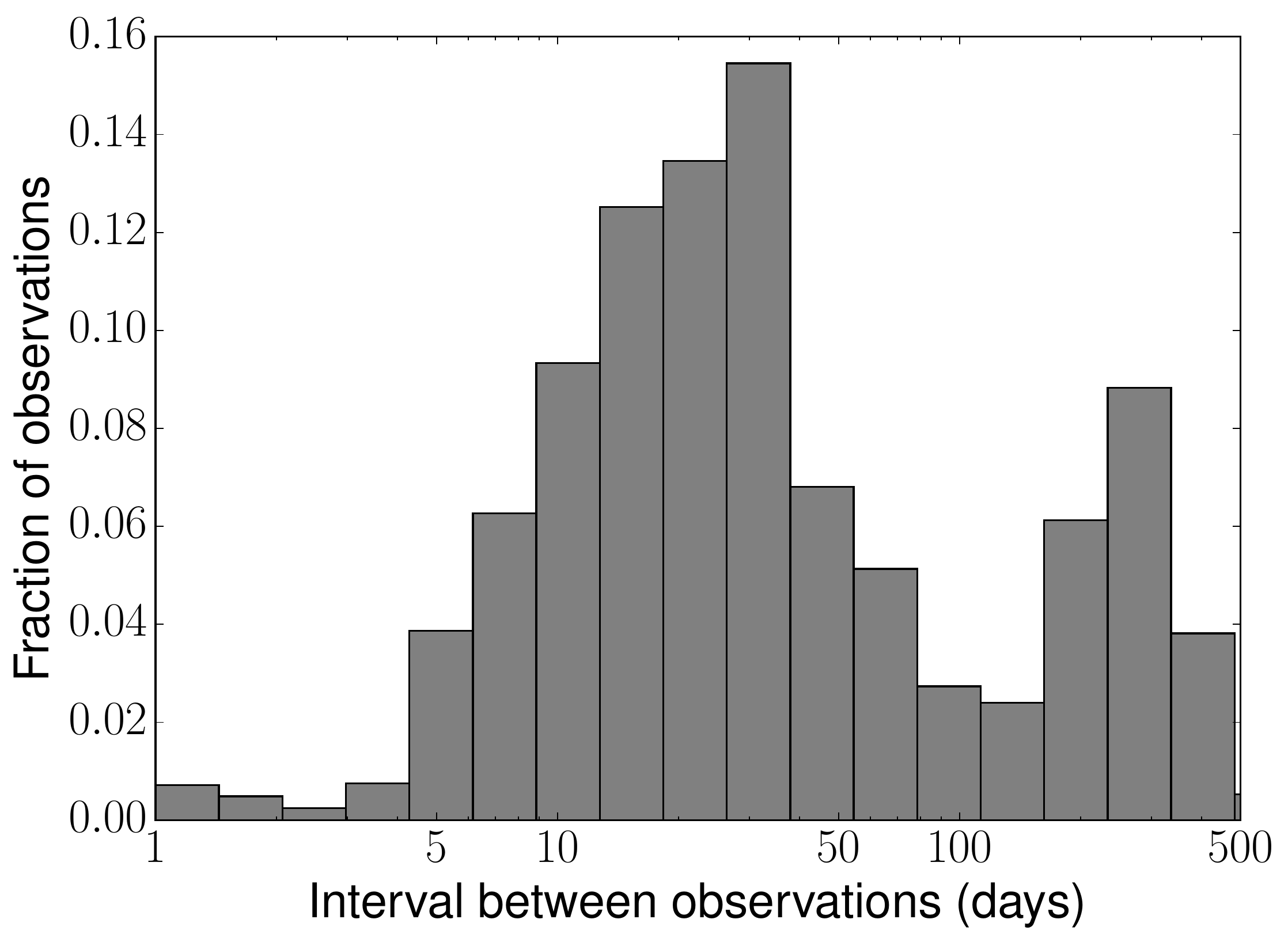}
  \caption{Distribution showing the interval, in days, between CRTS observations. The median interval was 28 days, 16\% of the intervals were less than 10 days, and 61\% were between 10 and 100 days.}
  \label{fig:hist_observationinterval}
\end{figure}

An initial, tentative classification is assigned based on the histogram of the light curve. A typical DN is predominantly in quiescence, so a histogram of its light curve will reflect this and have a higher peak in the fainter half of the magnitude range (see Figure \ref{fig:lc_standardDN}). Light curves that do not show these properties are typically of CVs that show extended high-states, such as novalikes and magnetic CVs, and are assigned the classification `non-DN'. This means that some DN with very high duty cycles ($>50\%$), and intermediate polars that show outbursts as well as high-states, can be incorrectly classified as `non-DN'. Although few eclipses are expected in the averaged data, CVs are flagged as potential eclipsers if the magnitude range exceeds $\Delta V=5$ mag, or the highest peak is within $V=0.4$ mag of the brightest point (which will be the case if there are no outbursts). The $\Delta V=5$ magnitude limit will miss the shallower eclipsers (e.g. grazing eclipses of the disc), but it is set to prevent the magnetic CVs and those with large orbital modulations from masquerading as eclipsers.

DN and potential eclipsers then undergo a second round of classification. In order to test the `DN' classification, the script traces and counts the outbursts, and then runs a number of checks to ensure that the outbursts are not high-states. Tracing the outbursts is necessary, as the CRTS light curve may be sampled up to four times per month (see Figure \ref{fig:hist_observationinterval}), and it is therefore insufficient to count every bright point as a separate outburst, as outbursts can last for more than a week. Assuming (temporarily) that the `DN' classification is correct, the apparent quiescent magnitude ($v_{\rm Q}$) is set equal to the magnitude of the highest peak of the histogram. Thereafter the script identifies the outbursts. Tracing through the light curve, an outburst `starts' at the first point that is $\Delta V=1.5$ mag brighter than $v_{\rm Q}$ and 1.5$\sigma$ above the scatter. The outburst is then traced until either (1) the lightcurve drops to within 1.5 mag of $v_{\rm Q}$, or (2) lasts more than 70 days, or (3) shows a turnover\footnote{A stage where the light curve shows a decrease, and subsequent increase, in brightness. As the light curve is averaged by day, and not sampled on consecutive days, a turnover is not expected within an outburst. Although some WZ Sge-type DN do show post superoutburst re-brightenings (e.g. \citealt{Kato2009,Nakata2013}), few turnovers are expected in the dataset due to the CRTS sampling cadence.}.

\begin{figure}
  \centering
    \includegraphics[width=8cm]{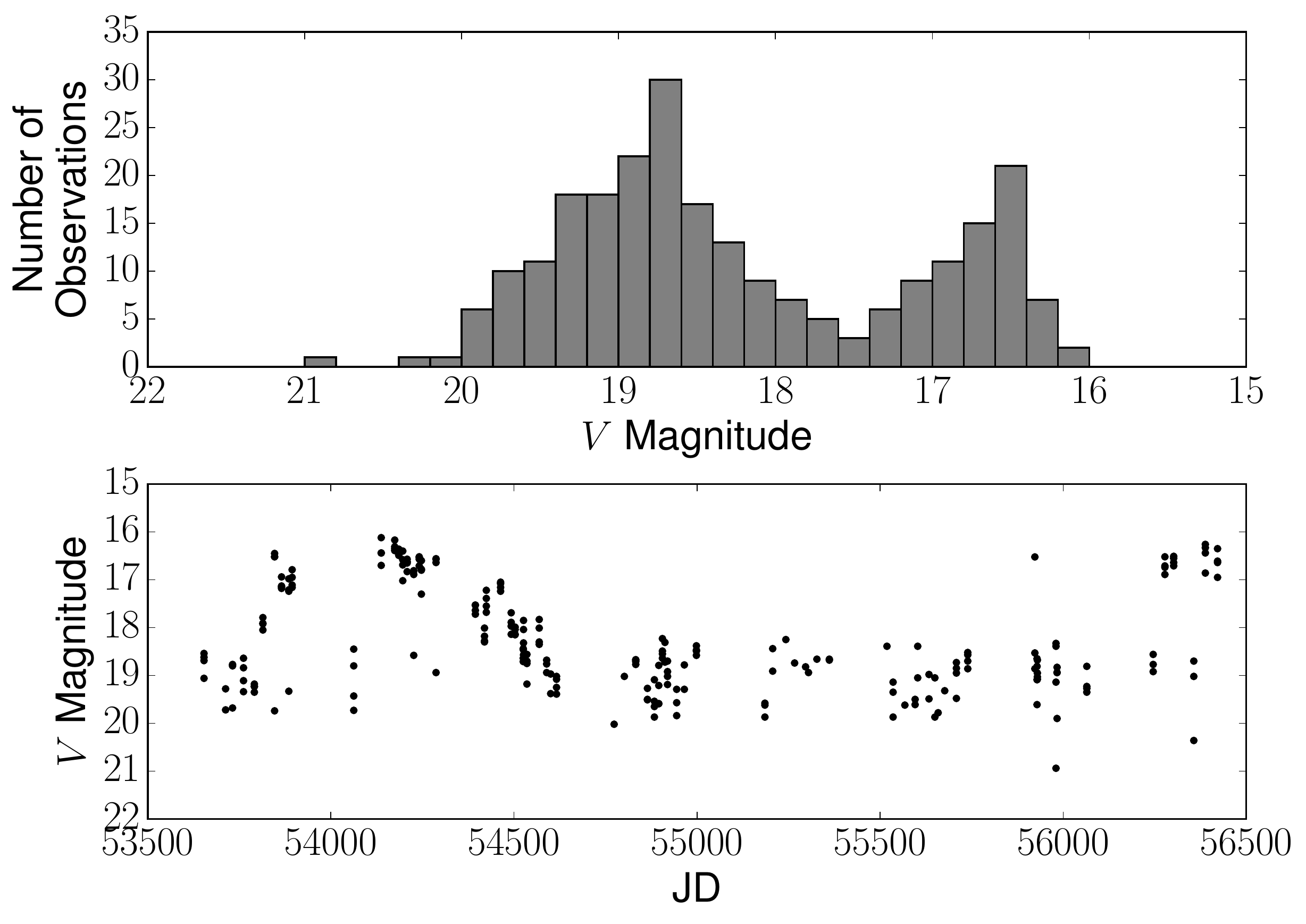}
    \caption{Example of a light curve that was designated as `non-DN' by the classification script because it shows high-states and low-states in the light curve (see Figure \ref{fig:code_flowchart}). Magnetic CVs typically show high-states in their optical light curves.}
  \label{fig:lc_highstates}
\end{figure}

With this procedure it is possible to mistake high-states for outbursts, so each potential outburst is tested. If an `outburst' shows any of the following properties, the CV classification is changed to `non-DN'. (1) If there is more than 1 turn-over per outburst, or (2) if an outburst lasts more than 100 days, or (3) consists of more than 8 light curve points, or (4) lasts more than 30 days, consists of more than 5 points and shows a scatter of $<0.8$ mag, then it is likely a high-state. Figure \ref{fig:lc_highstates} shows an example of a CV that was classified as a `DN' in the initial step, but was flagged as a `non-DN' in this step because high-states were found. Additionally, any CV that shows no outbursts is classified as `non-DN'.

Once the classification is complete, the script estimates the outburst properties for every CV classified as a DN. The brightest outburst point is taken as the apparent magnitude in outburst ($v_{\rm O}$), and $v_{\rm Q}$ is set equal to the magnitude of the highest histogram peak. As the CRTS did not necessarily detect the CV at the peak of outburst, $v_{\rm O}$ should be considered a faint-limit. The duty cycle is estimated as the fraction of time spent in outburst (number of days in outburst divided by the total number of days observed -- non-detections included\footnote{Note that image artifacts and saturation can cause a non-detection.}). The shortest time between two outbursts is taken as an upper-limit on the recurrence time, since the CRTS may have missed intervening outbursts due to the sampling cadence\footnote{A comparison of our recurrence time upper-limits to the recurrence-times listed in RKCat (19 DN in common) indicates that our upper-limit is approximately equal to the true recurrence time if 5 or more outbursts are observed in the light curve.}.

All the limits and criteria used in the script were defined in order to maximize the number of classifications and avoid misclassifications.

To determine the accuracy and efficiency with which the script classifies DN, we compared the classifications to those in the Catalogue of Cataclysmic Binaries, Low-Mass X-Ray Binaries and Related Objects \citep{Ritter2003}, hereafter refered to as RKCat. Of the 252 CRTS CVs with RKCat classifications, 243 had clear `DN' or `non-DN' classifications in RKCat. For this purpose, `DN' RKCat classifications include dwarf novae (DN, UG, ZC), SU UMa stars (SU, WZ), ER UMa stars (ER), or systems showing superhumps (NS,SH)\footnote{See http://www.mpa-garching.mpg.de/RKcat/ for further details on the classifications}. `Non-DN' RKCat classifications include novalikes (NL, SW, UX, VY), polars (AM, AS) and intermediate polar (IP, DQ)\footnote{In the case that a CV is classified as both `DN' and `non-DN' in RKCat, for example IP DN, it is given the classification `DN'}. For comparison purposes we assume that the RKCat classifications are all correct, but there may be DN with very long recurrence times that are misclassified as NL.

Of these 243 common CVs, 209 were classified as `DN' by the classification script and 28 were classified as `non-DN'. The accuracy of the `DN' sample is 95.7\% (200 out of 209 were also classified as `DN' in RKCat). The 9 incorrect DN were polars according to RKCat, and all had large amplitude, short-duration high-states (similar to DN outbursts) in their lightcurves\footnote{The classification for these 9 CVs has been corrected to `non-DN' in the Outburst Catalogue}. The efficiency with which the script finds DN is lower, as the accuracy of the `non-DN' classification was only 67.9\% (19 of the 28 were correctly classified as `non-DN' according to RKCat). There are a number of reasons for the lower efficiency. First, the light curve may not be sufficiently well-sampled to catch the outbursts. Second, the quiescent level may be undetectable. In this case, the script will mistake the outburst points for a quiescent level and count zero outbursts. This was the case for the majority of the DN mistaken for non-DN, as they were SU UMa stars with high amplitude outbursts and non-detectable quiescent levels. Currently the CRTS do not provide upper-limits for their non-detections, but in future CRTS data releases we will use the non-detections to correct this bias. Third (as mentioned previously), DN with very high duty cycles can also be mistaken for polars. The efficiency with which the code detects DN could be increased by relaxing the classification conditions, but it would decrease the accuracy. Relaxing the conditions under which a CV is classified as a `DN' would increase the efficiency, but it would also decrease the $\approx96$\% accuracy of the `DN' classifications.

\section{The Outburst Catalogue}\label{sec:catalogue}

\begin{figure}
  \centering
    \includegraphics[width=8cm]{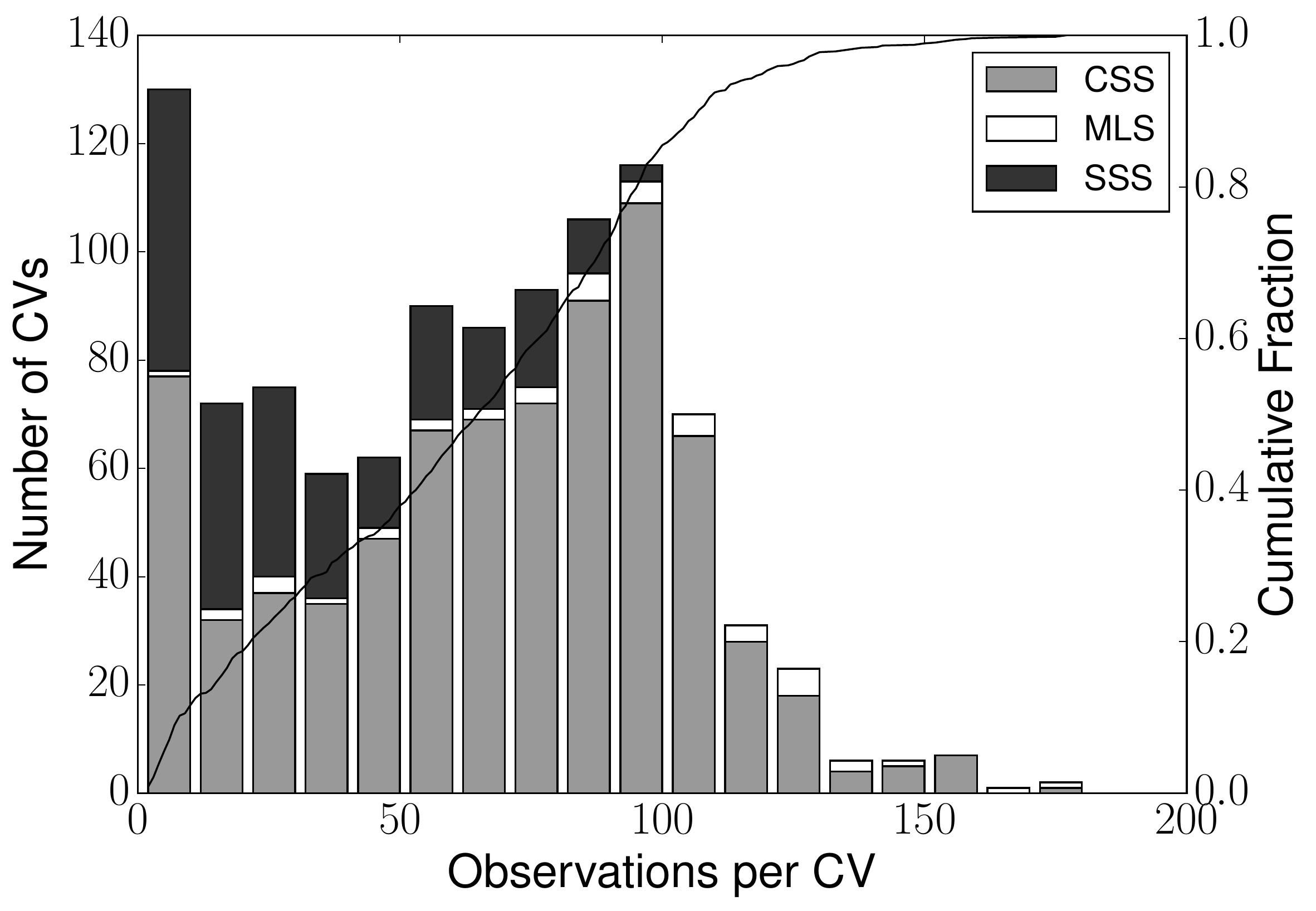}
    \caption{Distribution showing the number of CRTS observations per CV over the time-range used to make this catalogue, as well as the normalised cumulative distribution. The sampling cadence is uneven, and varies as a function of time and position on-sky. The variation between the sub-surveys is as a result of the different survey lengths and observing strategies (see \citealt{Drake2014}). Note that CVs with non-unique CRTS identifiers are only plotted once -- if a CV is detected in more than one sub-survey it will be plotted under its CSS identifier.}
  \label{fig:hist_numobservations}
\end{figure}

\begin{figure}
  \centering
    \includegraphics[width=8cm]{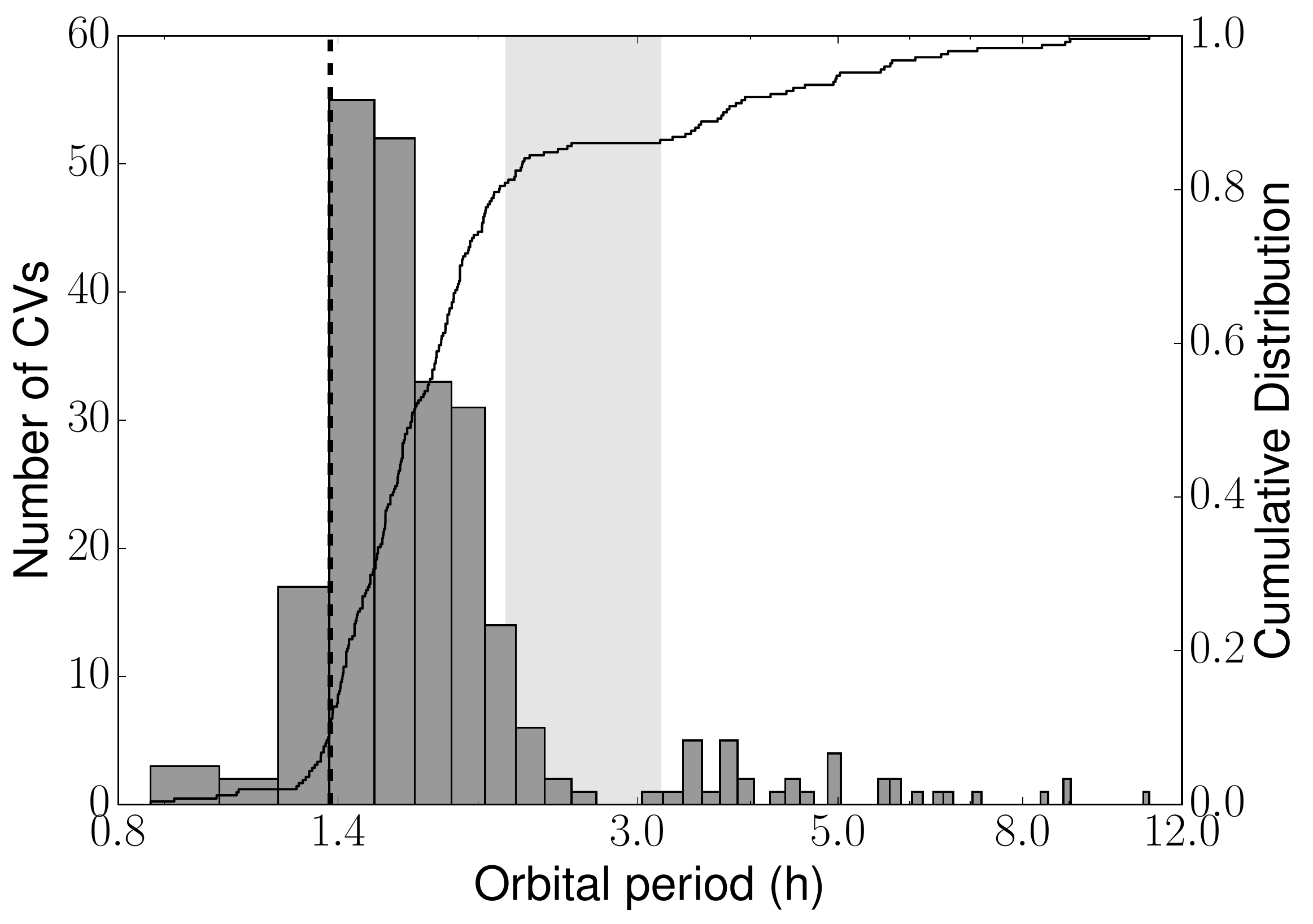}
    \caption{Orbital period distribution of the CRTS CVs with known $P_{\rm orb}$ in RKCat \citep{Ritter2003}. The cumulative distribution is indicated by the solid line, the shaded region indicates the period gap (with boundaries at 2.15 and 3.18 hours, \citealt{Knigge2006}) and the dotted line indicates the period minimum determined from the SDSS CVs ($\approx82$ min, \citealt{Gaensicke2009}). The nova X Serpentis ($P_{\rm orb}=35.52$ d, see \citealt{Thorstensen2000}) has been omitted for clarity.}
  \label{fig:hist_orbperiods}
\end{figure}

The Outburst Catalogue contains outburst properties (duty cycle, apparent $V$-band magnitudes in quiescence ($v_{\rm Q}$) and outburst ($v_{\rm O}$), and upper-limits on the recurrence time), system parameters, distance estimates, colour information, and X-ray counterparts where applicable, for 1031 CVs (of which 722 are DN), and 7 known AM CVn\footnote{Helium-rich, ultra-compact binary systems.}. This is the largest sample of estimates for these properties, which are seldom available, but often necessary when analysing particular systems or selecting objects for targetted observations/studies.

\begin{table*}
 \centering
 \begin{minipage}{140mm}
  \caption{Contents of the Outburst Catalogue}
  \begin{tabular}{llll}
    \hline
    Column Number & Column Description & Units & \# CVs with entry$^a$\\
    \hline
    1 & CRTS ID & -- & 1031\\
    2 & Alternative Names & -- & 1031\\
    3 & RA & deg & 1031\\
    4 & DEC & deg & 1031\\
    5 & CRTS sub-surveys & -- & 1031\\
    6 & Classification$^b$ & -- & 898\\
    7 & RKCat classification$^c$ & -- & 252\\
    8 & Spectrum$^d$ & -- & 242\\
    9 & $P_{\rm orb}$ & h & 143\\
    10 & $P_{\rm SH}$ & h & 179\\
    11 & $P_{\rm orb}$ from $P_{\rm SH}^e$ & h & 109\\
    12 & Inclination & deg & 17\\
    13 & \# CRTS Observations & -- & 1031\\
    14 & \# CRTS detections & -- & 1031\\
    15 & \# Outbursts$^b$ & -- & 715$^f$\\
    16 & Apparent outburst magnitude faint-limit$^b$ $\left(v_{\rm O}^{\rm lim,f}\right)$ & mag & 715$^f$\\
    17 & Apparent quiescent magnitude$^b$ ($v_{\rm Q}$) & mag & 614$^f$\\
    18 & Recurrence time upper-limit$^b$ $\left(t_{\rm recur}^{\rm lim,u}\right)$ & days & 570$^f$\\
    19 & Duty Cycle$^b$ & -- & 715$^f$\\
    20--24 & SDSS $u, g, r, i, z$ & mag & 94\\
    25--28 & WISE $w1, w2, w3, w4$ & mag & 332\\
    29--31 & 2MASS $J, H, K$ & mag & 147\\
    32--35 & UKIDSS $Y, J, H, K$ & mag & 39\\
    36 & Distance lower-limit$^g$ & pc & 88\\
    37 & Distance upper-limit$^h$ & pc & 206$^f$\\
    38 & Bright-limit on absolute quiescent magnitude$^i$ $\left(V_{\rm Q}^{\rm lim,b}\right)$ & mag & 71\\
    39 & Faint-limit on absolute quiescent magnitude$^j$ $\left(V_{\rm Q}^{\rm lim,f}\right)$ & mag & 197$^f$\\
    40--41 & ROSAT 0.1--2.4 keV count rate and error & counts/s & 35\\
    42--44 & Chandra 0.5-7 keV flux and upper- and lower-limits & mW/m$^2$ & 2\\
    45--46 & XMM 0.2--12 keV flux and error & mW/m$^2$ & 17\\
    \hline
  \multicolumn{4}{p{14cm}}{\textit{Notes:} The full catalogue is available online. $^a$Number of CVs that have an entry in the corresponding column (from a total of 1031 CVs -- the 7 known AM CVns are not included). $^b$Determined by the classification script (see Section \ref{sec:code}). $^c$ Classifications from RKCat \citep{Ritter2003} $^d$There are spectra for 242 CVs in \citet{Breedt2014}. $^e$Estimated using $P_{\rm orb}=0.9162(52)P_{\rm SH} + 5.39(52)$ (Equation 1 from \citealt{Gaensicke2009}, where periods are in minutes). $^f$Out of a total of 722 CVs classified as DN by the classification script. $^g$Derived by taking the apparent magnitude of the secondary as the WISE (or UKIDSS) $K$-band value, and estimating the absolute magnitude of the secondary from $P_{\rm orb}$ and the donor sequence from \citet{Knigge2011}. This method typically underestimates the true distance by a factor 1.75, as the secondary only contributes $\sim33$\% of the light in $K$-band (see \citealt{Knigge2006} for a discussion). $^h$Distance derived from $v_{\rm O}^{\rm lim,f}$ in column 16, and the absolute magnitude in outburst ($V_{\rm O}$) estimated from the $P_{\rm orb}$--$V_O$ relation \citep{Warner1987,Patterson2011}. This is then multiplied by a factor two to obtain a more robust upper-limit -- see Appendix \ref{sec:distance2_uncertainties}. $^i$Derived from column 36 and 17. $^j$Derived from column 37 and 17. \textit{Catalogue References}: 9--10, 12: RKCat, \citet{Ritter2003}. 20--24: Sloan Digital Sky Survey data release 8 \citep{Aihara2011}. 25--28: Wide-field Infrared Survey Explorer All-Sky data release \citep{Cutri2012}. 29--31: Two Micron All-Sky Survey All-Sky Catalogue of Point Sources \citep{Cutri2003,Skrutskie2006}. 32--35: United Kingdom Infra-red Deep Sky Survey data release 10 \citep{Lawrence2007}. 40--41: R\"{o}ntgensatellit All-sky survey faint, and bright, source catalogues \citep{Voges1999,Voges2000}. 42--44: Chandra Source Catalogue 1.1 \citep{Evans2010}. 45--46: 3XMM-DR4 catalogue \citep{Watson2009,Rosen2015}. A cone search, of radius 1.2$''$ (SDSS), 3$''$ (WISE), 2$''$ (2MASS), 1.2$''$ (UKIDSS), 10$''$ (ROSAT), 2$''$ (Chandra) and 4$''$ (XMM), was used to match catalogues, and only unique matches were recorded.}\\
  \end{tabular}\label{tbl:catalogue}
 \end{minipage}
\end{table*}

\begin{figure*}
  \centering
    \includegraphics[width=15cm]{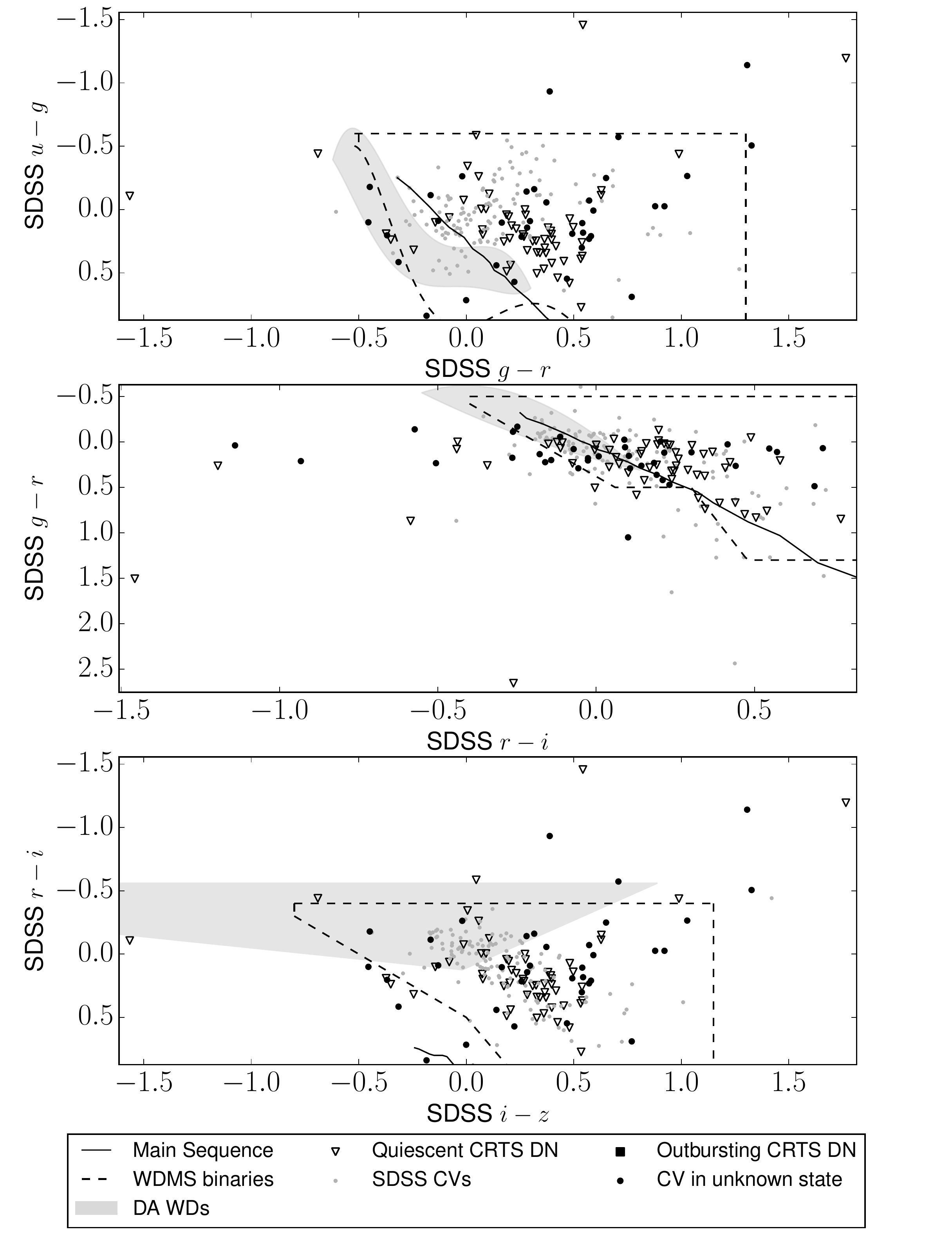}
    \caption{Colour-colour distributions for the CRTS CVs, using colour information from SDSS DR8 \citep{Aihara2011}. Where possible, we have distinguished between the colours during outburst and quiescence (see text). Those identified as `non-DN' by our script are labelled as `CV in unknown state'. Selection boxes for other classes of objects are shown for reference. SDSS CVs: Colour selected CVs from the SDSS \citep{Gaensicke2009}. Main Sequence: Star colours from models optimized through comparison to the Praesepe cluster \citep{Kraus2007}. DA WDs: Observational colour-colour cuts for WDs with Hydrogen-rich atmospheres based on SDSS DR7 \citep{Girven2011}. WDMS binaries: White Dwarf Main-sequence binaries selected from SDSS DR8 \citep{Rebassa-Mansergas2013}. Only colours for those CVs with a clean photometry flag in the SDSS were recorded in the catalogue. The extreme outliers are most likely to be as a result of photometric errors.}
  \label{fig:colour_colour}
\end{figure*}

The data-set used to make this catalogue is the Catalina Surveys Data Release 2 (CSDR2), covering the dates 2005-04-04 to 2013-10-31 (CSS), 2005-06-12 to 2014-01-23 (MLS) and 2005-04-19 to 2013-07-22 (SSS). A histogram of the number of observations per CV in this dataset is shown in Figure \ref{fig:hist_numobservations}, and as mentioned previously, Figure \ref{fig:hist_observationinterval} shows the length of the intervals between CRTS observations. Table \ref{tbl:catalogue} describes the columns in the Outburst Catalogue, which can be accessed online and at the Strasbourg astronomical Data Center\footnote{http://cdsarc.u-strasbg.fr/} (CDS).

\section{Analysis and Discussion}\label{sec:discussion}

In all subsequent analysis, the 7 known AM CVns have been excluded, as these systems have a different evolutionary path and outburst characteristics to CVs \citep[e.g.][]{Levitan2015}. 

\subsection{The Orbital Period Distribution}\label{sec:orbperiod_dist}

Figure \ref{fig:hist_orbperiods} shows the orbital period distribution of the CRTS CVs. The sample shows a clear peak that is consistent with the period spike determined from the SDSS CVs (80--86 mins, \citealt{Gaensicke2009}). There are 14 CVs with orbital periods shorter than 80 mins. According to RKCat, 5 are WZ Sge-type CVs and the remainder are SU UMa-type CVs. See \citet{Breedt2014} for a detailed discussion of these systems.  

The figure also shows that the CRTS is detecting a large population of CVs with $P_{\rm orb}$ below the period gap. This has been noted previously in the literature (e.g. \citealt{Woudt2012, Thorstensen2012, Drake2014}). Since it is expected that the short orbital-period systems should dominate the population and that the fraction of DN above the period gap should be smaller (e.g. \citealt{Knigge2011}), the deeper surveys are now revealing more of the intrinsic population (e.g. \citealt{Breedt2014,Gaensicke2009}).

\subsection{Colour-colour plots}\label{sec:colour-colour}

\begin{figure}
  \centering
    \includegraphics[width=8cm]{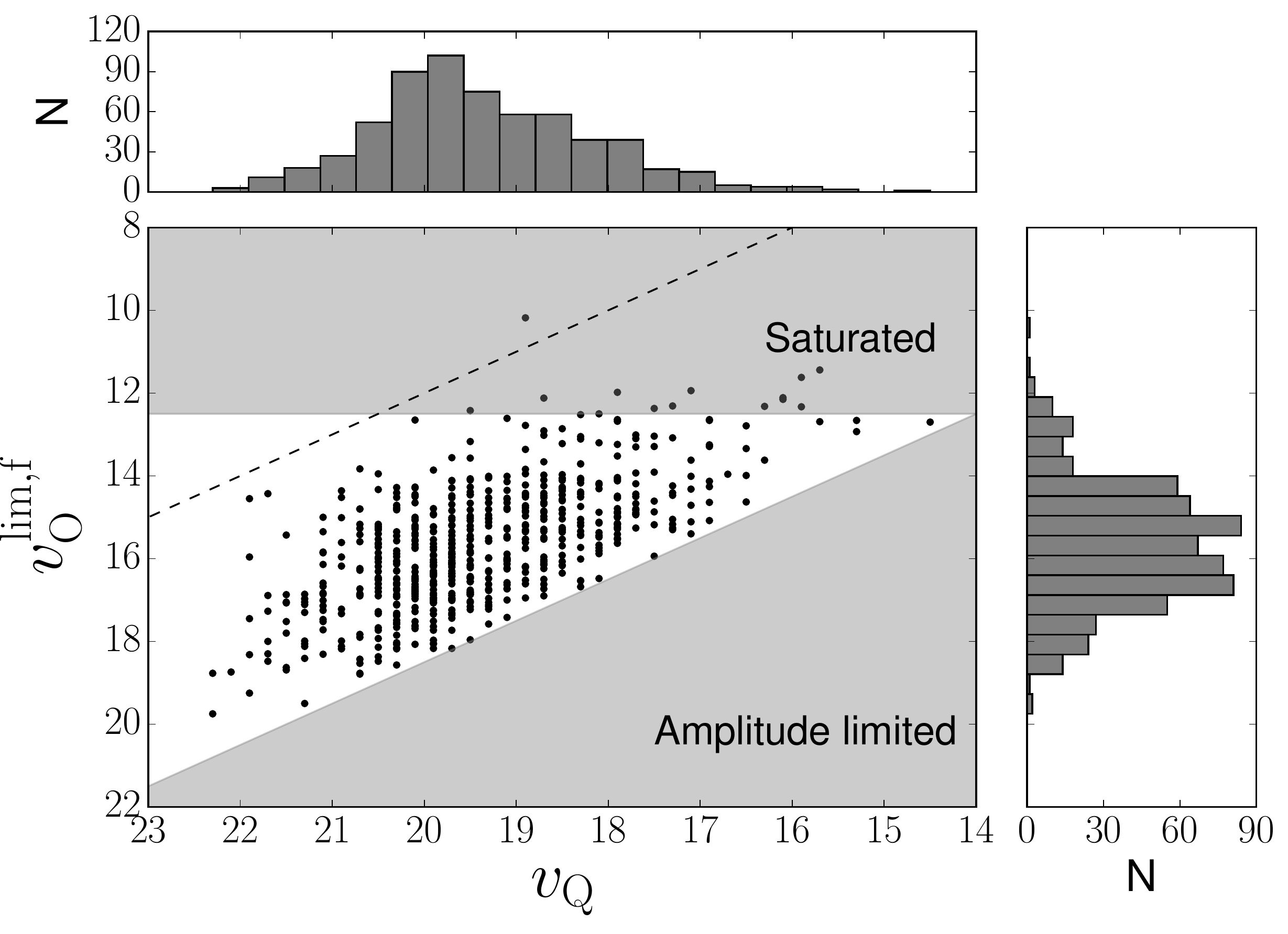}
    \caption{Distribution of the apparent outburst and quiescent magnitudes, with selection effects indicated. The upper shaded region indicates the range for which the object is saturated in the CRTS image and flagged out as an artifact. Note that the exact saturation level varies according to seeing conditions and sub-survey. The lower shaded region indicates the range where the outburst amplitude is less than 1.5 mag, and hence not considered an outburst by the classification code. The official detection limits for the CRTS are quoted as $V=21.5$ mag (MLS), $V=19.5$ mag (CSS) and $V=19.0$ (SSS), although it is clear from the figure that it can see much deeper when the conditions are good -- the faintest detections in the database are at 22.9 (MLS), 23 (CSS) and 22.2 (SSS). The deep co-added reference images for each field are typically at $V=21.5$ (CSS), $V=22.5$ (MLS) and $V=20$ (SSS). An outburst amplitude of $\Delta V=8$ mag is indicated by the dotted line.}
  \label{fig:outburst_vs_quiescence}
\end{figure}

Colour-colour plots for the CVs with SDSS DR8 photometry are shown in Figure \ref{fig:colour_colour}. Where possible, we have distinguished between DN that were observed in quiescence and DN observed in outburst by the SDSS. \citet{Drake2014} determined that the CRTS $V$-band and SDSS $i$-band correspond most closely, with an average difference of --0.01 mag with $\sigma=0.33$ mag. Consequently we determined a DN to be in outburst if $i\leq v_{\rm O}$ or $v_{\rm Q}-i\geq 1.5$ mag, or in quiescence otherwise. Only 1 DN was found to be in outburst at the time of the SDSS observations (CSS080409:174714+150048 had $v_{\rm Q}-i=2.5$ mag), but there are likely to be more in outburst, or on the rise to outburst, that do not meet our cautious definition of what constitutes an outburst. There is however, a CRTS bias against detecting DN that were in outburst in the SDSS, as the SDSS magnitude is used as a baseline for determining variability (see \citealt{Wils2010}). CVs that were not classified as DN are plotted as in an unknown state. The SDSS colour-selected CVs \citep{Gaensicke2009}, and the variability-selected CRTS CVs do have the same locus in the plots, but the CRTS CVs cover a larger range of colours.

\subsection{Outburst Properties}\label{sec:dutycycle}

Out of the 1031 CVs, 722 were classified as DN by the classification code. We now characterise the outburst properties of this sample and discuss the selection effects.

\begin{figure}
  \centering
    \includegraphics[width=8cm]{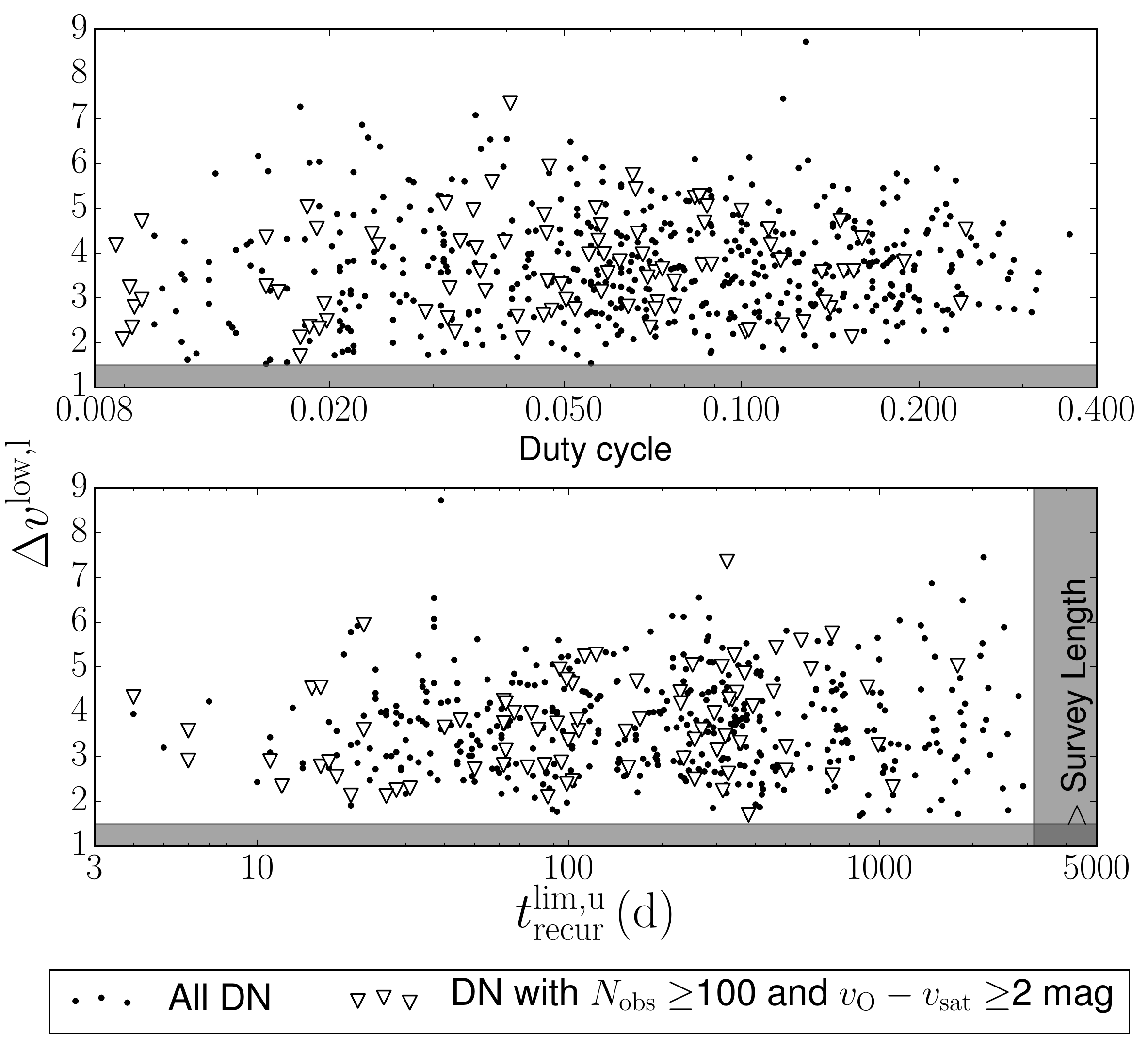}
    \caption{Distribution of the outburst amplitude lower-limit with duty cycle (top panel), and the upper-limit on the recurrence time (lower panel). The shaded regions indicate where the range exceeds the survey length, and where the outburst amplitude is smaller than $1.5$ mag. The well-sampled DN (those with more than 100 CRTS observations that did not have $v_{\rm O}^{\rm lim,f}$ within 2 mag of the saturation limit) are indicated to show that the observational effects are not masking correlations between the properties. The density fluctuations in the recurrence time limit are as a result of CRTS observing cycles due to seasonal variations.}
  \label{fig:amp_vs_dcandrecurtime}
\end{figure}

Figure \ref{fig:outburst_vs_quiescence} shows the distribution of quiescent and outburst apparent magnitudes. The median $v_{\rm O}^{\rm lim,f}=15.7\,$mag, $v_Q=19.5\,$mag, and $\Delta V=3.6\,$mag which corresponds to an lower-limit for the outburst amplitude of $\Delta V^{\rm lim,l}=2.8$ mag. The largest DN outburst amplitude is expected to be $\Delta V\approx8$ mag, but much of the $\Delta V>8$ mag phase space falls above the CRTS saturation limit. The CRTS and classification script variability criteria also limit the outburst amplitude to $\Delta V\geq1.5$ mag.

\begin{figure}
  \centering
    \includegraphics[width=8cm]{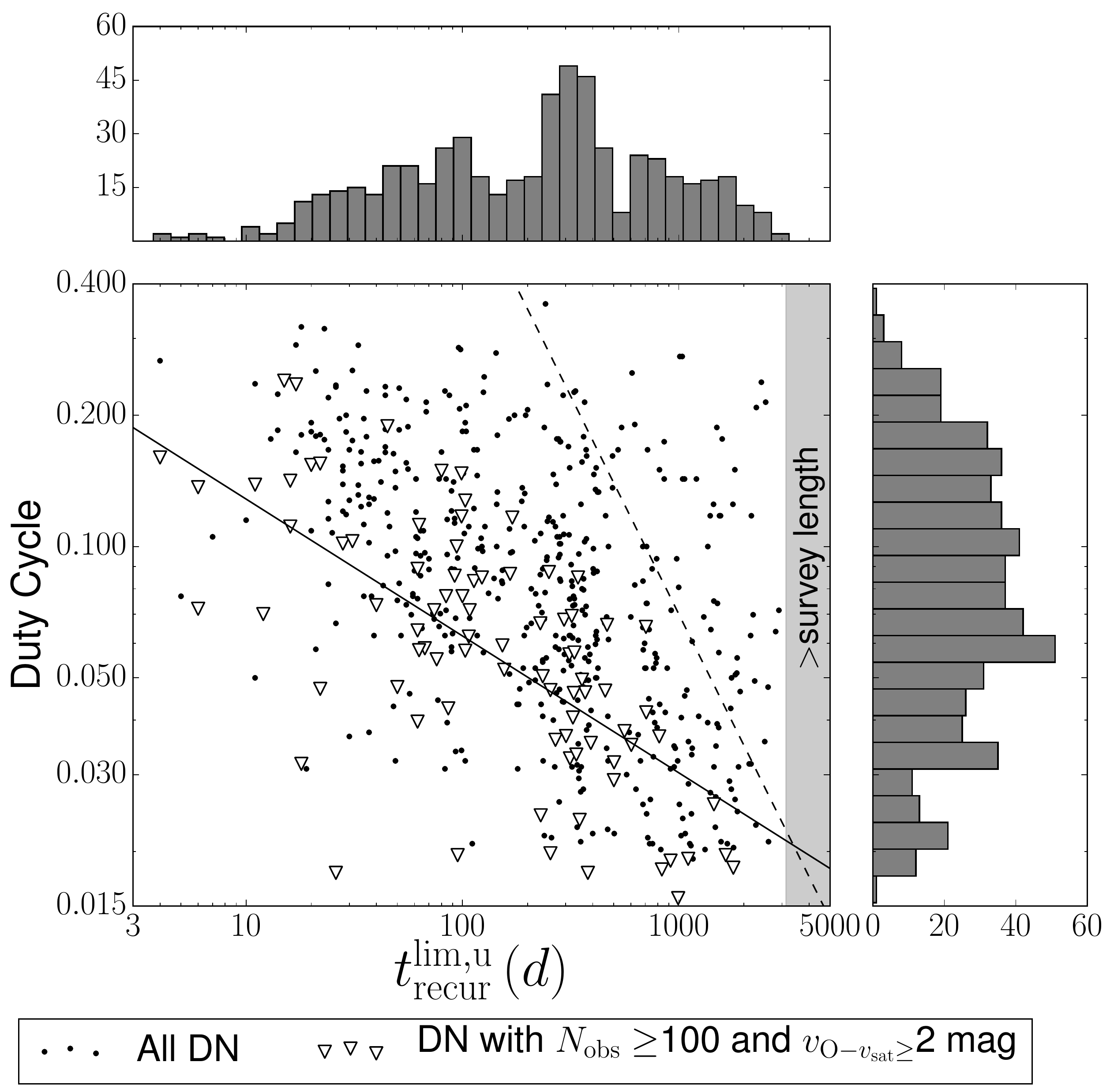}
    \caption{Distribution of the duty cycle and the upper-limit on the recurrence time, showing a strong correlation between the duty cycle and $t_{\rm recur}^{\rm lim,u}$ (solid line). The shaded region indicates the range where the recurrence-time exceeds the survey length. Outbursts are limited to 100 days by the classification script in order to distinguish between high- and low-states. This restriction is indicated by the dashed line. Poor sampling can make the duty cycle appear larger if a large fraction of outburst points are observed.}
  \label{fig:duty_cycle}
\end{figure}

The sample does not show a correlation between $\Delta V^{\rm lim,l}$, and the duty cycle (see Figure \ref{fig:amp_vs_dcandrecurtime}). This is still the case if only the well-sampled DN are considered, so it is unlikely that the scatter is purely a result of the sampling and saturation limit. The CRTS is biased towards detecting the high duty cycle DN -- as pointed out by \citet{Breedt2014}, they have in fact discovered most of the high duty cycle DN, but are still detecting low duty-cycle DN.

In the lower panel of the figure, there are indications that shorter recurrence times have smaller outburst amplitudes. A Spearman rank-order correlation test on the well-sampled points gives $\rho=0.26$ and $p=1.9\%$. This means that the sample shows a weak positive correlation\footnote{$\rho=1$ (or $\rho=-1$) indicates a perfect monotonic correlation with a positive (or negative) trend.} and the probability ($p$) that the null hypothesis that two uncorrelated datasets would produce this $\rho$ value is 1.9\%.

Figure \ref{fig:duty_cycle} shows a correlation between the duty cycle ($dc$) and $t_{\rm recur}^{\rm lim,u}$. If only the well-sampled DN are considered, then they related according to 
\begin{equation}
 \log (dc)=-0.31(\pm0.04)\log (t_{\rm recur}^{\rm lim,u})-0.58(\pm0.09),
\end{equation}
and the duty cycle is lower for DN with longer outburst recurrence times (the outburst duration does not increase in proportion to the recurrence time). The Spearman rank-order correlation coefficients are $\rho=-0.6646$ and $p=5.4\times10^{-12}$, so the sample shows a strong negative correlation. Note however, that there is a selection bias against short recurrence-time, low duty cycle DN, due to the CRTS cadence and the ability of the classification script to distinguish between outbursts and high-states.

The median duty cycle and $t_{\rm recur}^{\rm lim,u}$ are 5.8\% and 138 days respectively for the well-sampled DN. For the overall population the median values are 8\% and 250 days. There are, however, several selection effects that influence this distribution. A DN with a high duty cycle is more likely to be detected and classified as a CV by the CRTS as it is highly variable. Only those objects that appear bright with respect to the co-adds are flagged as transients, but a high duty cycle is unlikely to reduce the detection probability -- as a DN would need to have an extremely high duty cycle to be in outburst in the CRTS reference image (which is composed of $\approx20$ coadded images). Additionally, the duty cycle would need to exceed 50\% to be incorrectly classified by the classification script\footnote{For a duty cycle exceeding 50\%, the majority of the light curve points would be in outburst and consequently the highest histogram peak would be in the brighter half of the magnitude range -- see Section \ref{sec:code}}. In contrast, the classification script introduces a bias against high duty cycles for the poorly sampled light curves. If the quiescent state between outbursts is not observed, separate outbursts can be mistaken for high-states, and the CV classified as `non-DN'. The recurrence time limit distribution also shows clear density fluctuations, which are produced by seasonal cycles in the CRTS observations. For example, fewer DN with a recurrence time of a year will be detected, because a fraction of them will have outbursts during the time when they are not observed by the CRTS.

\begin{figure*}
  \centering
    \includegraphics[width=\textwidth]{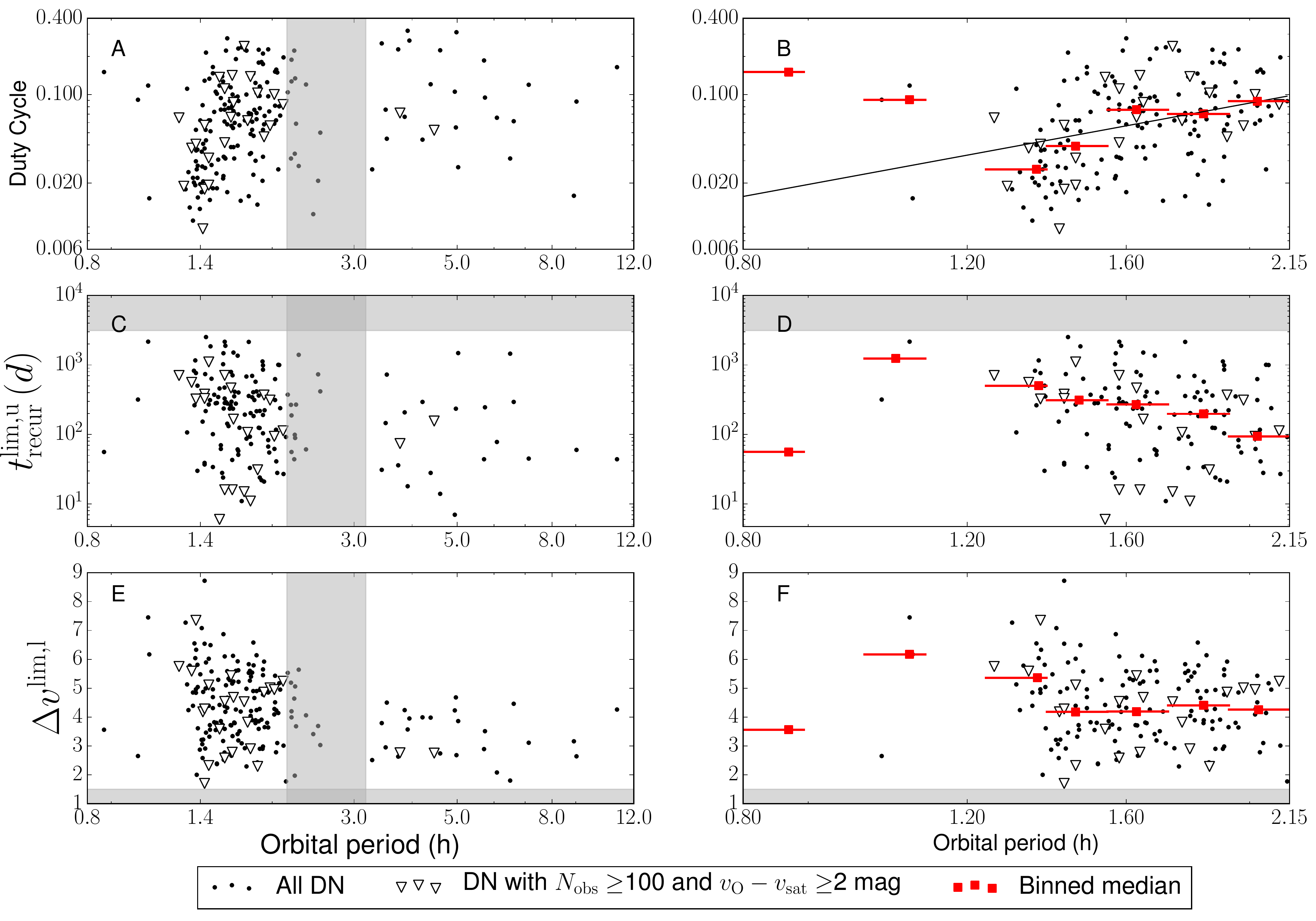}
    \caption{The dependence of the outburst properties (duty cycle, upper-limit on the recurrence time, and lower-limit on the outburst amplitude) on the orbital period. The panels on the left show the full $P_{\rm orb}$ range, while the panels on the right show the range $P_{\rm orb}<2.15$ h. Red squares indicate the median value within a given logarithmic period bin (the size of which is indicated by the error bars), to show the trend. The shaded regions indicate the period gap and survey limits. The nova X Serpentis ($P_{\rm orb}=35.52$ d, see \citealt{Thorstensen2000}) has been omitted for clarity.}
  \label{fig:dcvsporb_rtvsporb}
\end{figure*}

Outburst properties such as the amplitude, recurrence time, and duty cycle depend on the mass-transfer rate and the orbital period (as $P_{\rm orb}$ determines the size of the accretion disc). Figure \ref{fig:dcvsporb_rtvsporb} shows the distributions of these properties with $P_{\rm orb}$. Above the period gap, the angular momentum loss mechanism is magnetic braking, whereas below the gap it is predominantly gravitational radiation (see \citealt{Knigge2011}). These two regimes consequently have different mass-transfer rates, so we treat them separately. Below the period gap, the duty cycle shows a significant positive trend with $P_{\rm orb}$ (Spearman rank-order correlation coefficients are $\rho=0.41$ and $p=4\times10^{-8}$). A linear-least squares fit give a relationship of
\begin{equation}
 \log (dc)=1.9(\pm0.4)\log(P_{\rm orb})-1.63(\pm0.08)\, ,
 \label{eq:dc_porb_relation}
\end{equation}
where $P_{\rm orb}$ is in hours. Although $t_{\rm recur}^{\rm lim,u}$ shows a significant negative trend with $P_{\rm orb}$ (Spearman rank-order correlation coefficients: $\rho=-0.22$ and $p=0.01$), $\Delta V^{\rm lim,l}$ does not.

\begin{figure}
  \centering
    \includegraphics[width=8cm]{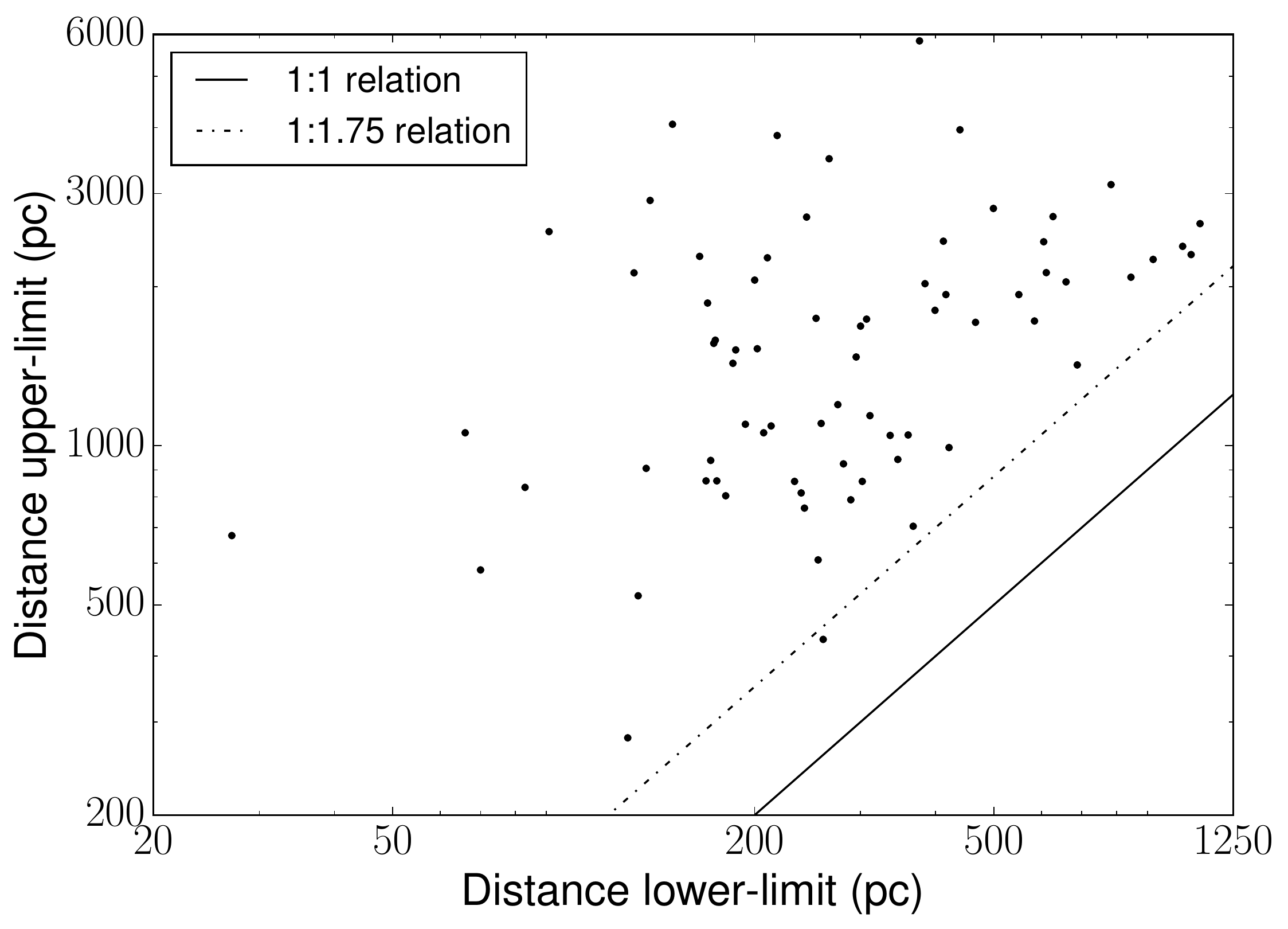}
    \caption{Distance range for the sample of DN with known orbital periods and $K$-band magnitudes. The upper-limit is determined using the $P_{\rm orb}$--$M_{V_{\rm max}}$ relation, and the lower-limit is determined using the the 2MASS (or UKIDSS) $K$-band magnitude -- see text for further details. For reference the 1:1, and 1:1.75, line are plotted. Note that the functional form of the relationship between the two distance limits would still be the same if the distance estimates were plotted, as the limits and estimates differ by a constant (see text for details).}
  \label{fig:dist1vs2}
\end{figure}

\begin{figure}
  \centering
    \includegraphics[width=8cm]{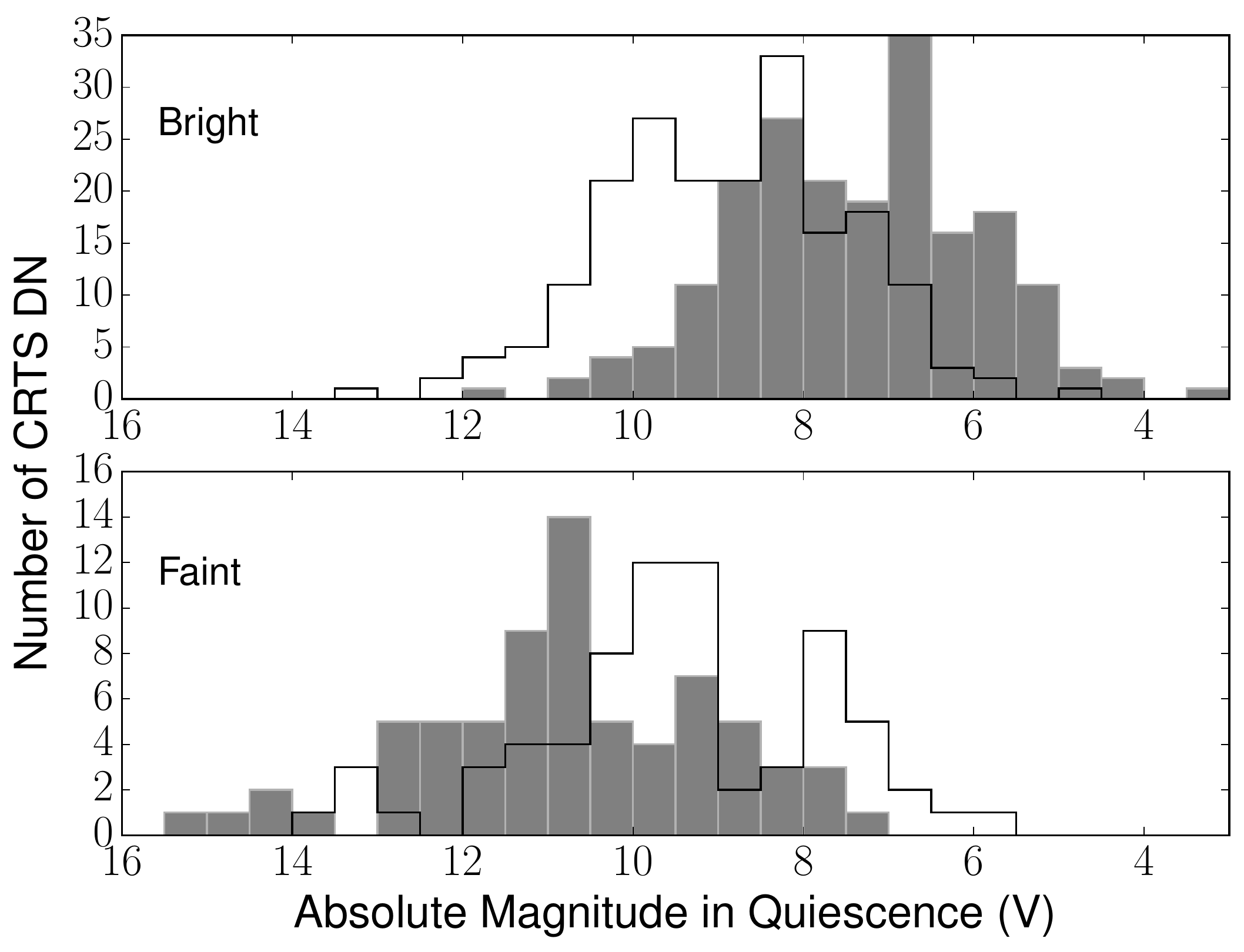}
    \caption{Distribution of the quiescent absolute magnitudes for the DN with distance limits. Top panel: Bright limit for the quiescent absolute magnitude ($V_{\rm Q}^{\rm lim,b}$). The solid line gives the equivalent estimate ($V_{\rm Q}^{b}$). Bottom panel: Faint limit for the quiescent absolute magnitude ($V_{\rm Q}^{\rm lim,f}$). The solid line gives the equivalent estimate ($V_{\rm Q}^{f}$).}
  \label{fig:hist_quiescence}
\end{figure}

\citet{Britt2015} found a relationship between the duty cycle and X-ray luminosities of DN. Unfortunately, given the uncertainty on our distance estimates, it is not possible to tell if this sample showed the same relationship.

\subsection{Distance and absolute magnitude estimates}\label{sec:distances_and_absmags}

\begin{figure}
  \centering
    \includegraphics[width=8cm]{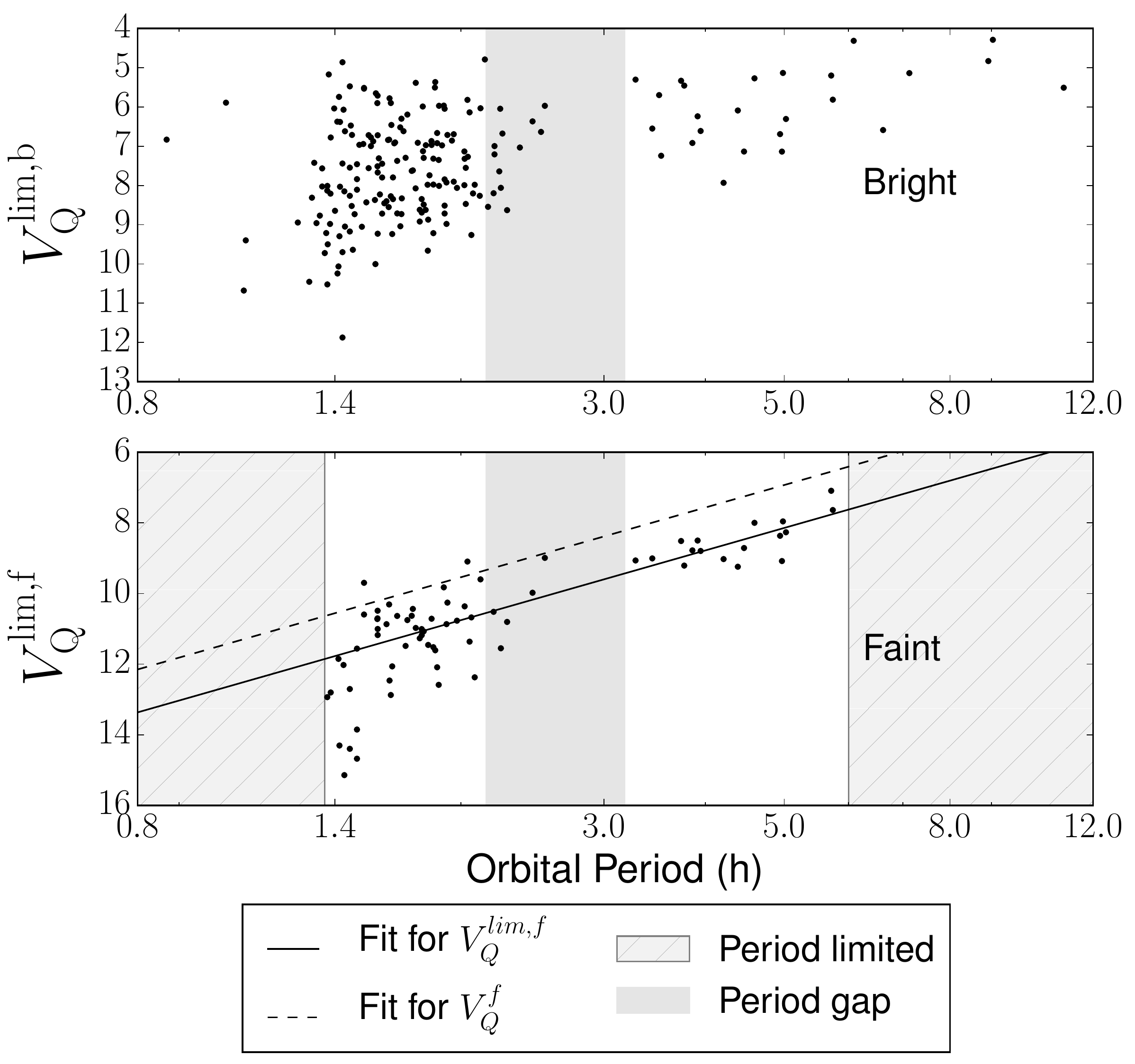}
    \caption{Variation of the quiescent absolute magnitudes with $P_{\rm orb}$. See text for a description of the symbols. The period gap is indicated by the gray region in both panels. Top panel: $V_{\rm Q}^{\rm lim,b}$ is directly derived from $P_{\rm orb}$, so this is plotted purely for comparison purposes. Bottom panel: The solid and dashed lines indicate the linear-least squares fit to $V_{\rm Q}^{\rm lim,f}$, and $V_{\rm Q}^{f}$ respectively. The range of orbital periods for which this distance determination method is not appropriate is indicated by the hatched region.}
  \label{fig:absmag_period}
\end{figure}

As described in detail in Table \ref{tbl:catalogue} and Appendix \ref{sec:distance2_uncertainties} (but repeated here briefly for clarity), two methods were used to determine distance limits. The upper-limit is derived by multiplying the distance estimate from the $P_{\rm orb}-V_{\rm max}$ relation \citep{Warner1987,Patterson2011} by a factor two, in order to compensate for the lack of known orbital inclinations. The lower-limit is derived using the 2MASS (or UKIDSS) $K$-band magnitudes and an estimate of the absolute magnitude of the secondary from the donor-sequence from \citet{Knigge2011}.

The distance limits for the DN in this sample are shown in Figure \ref{fig:dist1vs2}. The maximum distance of this sample appears to be less than 6000 pc. Consider, however, that in order to determine $v_{\rm O}^{\rm lim,l}$ and hence the distance, $v_{\rm Q}$ needs to be brighter than the CRTS detection limit. This excludes the (fainter) DN that the CRTS detected only in outburst. Additionally, only those DN with distance lower-limits  are shown, so the detection limits of 2MASS and UKIDSS are imposed on the plot as well. The CRTS is consequently probing a larger volume of DN than indicated by the distance limits in this catalogue.

\begin{figure}
  \centering
    \includegraphics[width=8cm]{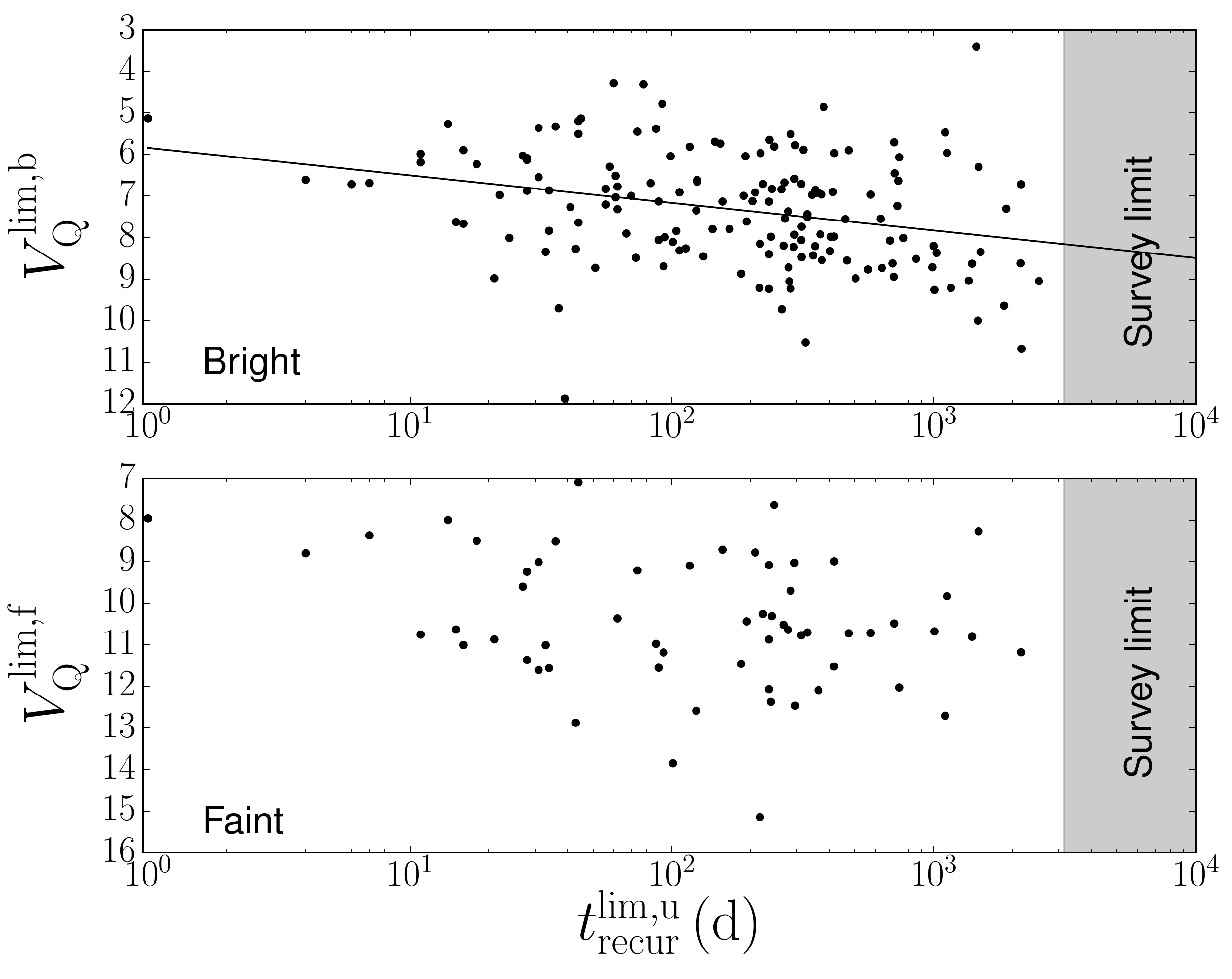}
    \caption{Distribution of the bright- and faint-limit of the absolute magnitude in quiescence with respect to the upper-limit on the outburst recurrence time. The solid line indicates the linear-least squares fit.}
  \label{fig:absmag_quiescence_vs_recurrencetime}
\end{figure}

Figure \ref{fig:dist1vs2} also shows a scarcity of CVs at short distances ($<500$ pc). The distance lower-limit is expected to underestimate the true distance by a factor 1.75, as the $K$-band magnitude is expected to contribute $\sim33\%$ of the light (see Table \ref{tbl:catalogue}). Nearby CVs are likely to be bright and saturated in outburst in the CRTS images, and consequently flagged as image artifacts, so the CRTS DN population will not be complete at small distances. 

The distribution for the quiescent absolute magnitude ($V_{\rm Q}$) limits derived from the distance limits and $v_{\rm Q}$ are shown in Figure \ref{fig:hist_quiescence} -- please see the caption for a description of the $V_{\rm Q}$ symbols. For the bright and faint limits, the median is $V_{\rm Q}^{\rm lim,b}=7.3$ mag and $V_{\rm Q}^{\rm lim,f}=10.7$ mag respectively. Using the distance estimates (instead of the limits) to calculate $V_{\rm Q}$, produces median values of $V_{\rm Q}^{b}=8.9$ mag and $V_{\rm Q}^{f}=9.5$ mag respectively, which are in good agreement.

The faintest limits in the distribution ($V_{\rm Q}^{\rm lim,f}\gtrsim14$) are not physical, as the temperature of the WD would imply a cooling time approaching the Hubble time. As the only quantity determined from the light curves for this estimate is $v_{\rm Q}$, poor sampling cannot explain this discrepancy. It is likely that the distances for these CVs were underestimated by more than the standard 1.75 expected from this method, which would imply that the donors may contribute less than 30\% of the light even in $K$-band. All four of the DN with $V_{\rm Q}^{\rm lim,f}\gtrsim14$, have short orbital periods (in the range 1.41--1.49 h), but are classified as SU UMa type CVs in RKCat, so it is unlikely that these are extremely evolved, ultra-faint donor stars.

\begin{figure}
  \centering
    \includegraphics[width=8cm]{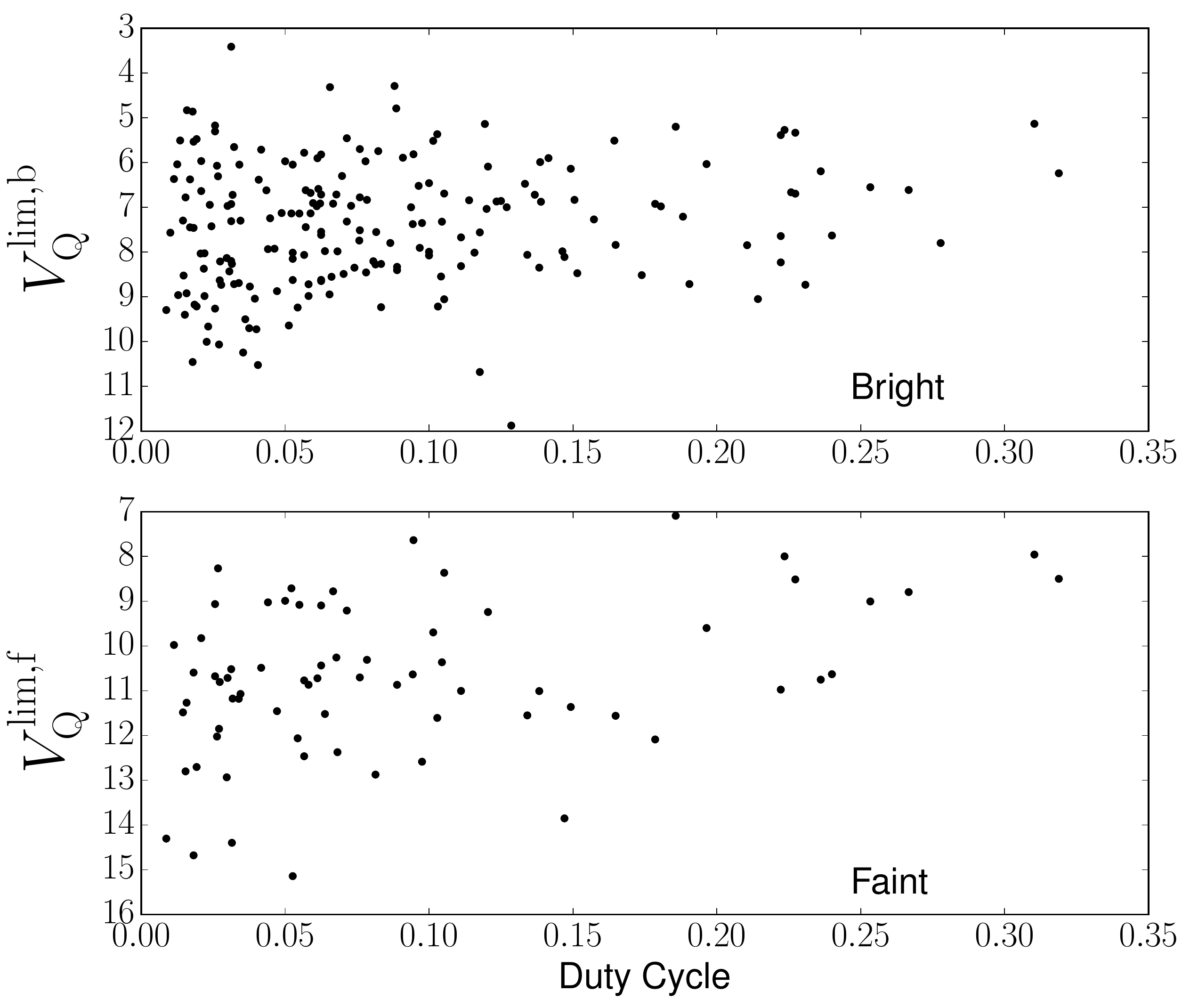}
    \caption{Distribution of the bright- and faint-limit of the absolute magnitude in quiescence as a function of the duty cycle.}
  \label{fig:absmag_quiescence_vs_dutycycle}
\end{figure}

Figure \ref{fig:absmag_period} shows a dependence of $V_{\rm Q}^{\rm lim,f}$ on $P_{\rm orb}$ of the form
\begin{equation}
 V_{\rm Q}^{\rm lim,f}=-6.6(\pm0.5)\times\log (P_{\rm orb})+12.7(\pm0.2)\, ,
\end{equation}
where $P_{\rm orb}$ is in hours. If the estimate $V_{\rm Q}^{f}$ is considered instead, the relationship is 
\begin{equation}
 V_{\rm Q}^{\rm f}=-6.6(\pm0.5)\times\log (P_{\rm orb})+12.7(\pm0.2)\,.
\end{equation}
In both cases, the unphysical points for which $V_{\rm Q}^{\rm lim,f}\gtrsim14$ have been excluded from the fit. $V_{\rm Q}^{\rm lim,b}$ was calculated using $P_{\rm orb}$ so it will show a dependence on $P_{\rm orb}$ -- the trend, however, is in the same direction as that of $V_{\rm Q}^{\rm f}$. The quiescent absolute magnitude is thus fainter at shorter orbital periods.

In Figure \ref{fig:absmag_quiescence_vs_recurrencetime}, $V_{\rm Q}^{\rm lim,b}$ shows a correlation (Spearman rank-order correlation coefficients are $\rho=0.32$ and $p=4.9\times10^{-5}$) with $t_{\rm recur}^{\rm lim,u}$ of the form
\begin{equation}
 V_{\rm Q}^{\rm lim,b}=0.64(\pm0.17)\log (t_{\rm recur}^{\rm lim,u})+5.89(\pm0.38)\, ,
\end{equation}
indicating that the systems with longer recurrence times are generally fainter. There were two DN with recurrence times of one day. In both cases the large scatter in the quiescent magnitude led the classification script to erroneously split one outburst into two, so these are artifacts. $V_{\rm Q}^{\rm lim,f}$ does not show a significant trend with $t_{\rm recur}^{\rm lim,u}$.    
 
Figure \ref{fig:absmag_quiescence_vs_dutycycle} shows that the limits for the quiescent absolute magnitude were correlated with the duty cycle. The Spearman rank-order correlation coefficients are $\rho=-0.32$ and $p=0.0066$ for the faint-limit $V_{\rm Q}^{\rm lim,f}$. The significance of the correlation for the bright-limit $V_{\rm Q}^{\rm lim,b}$ is lower, as the coefficients are $\rho=-0.16$ and $p=0.022$. As the estimates for the quiescent absolute magnitude differ from the limits by a constant factor (of less than 2), the functional form of the relation will be the same. This indicates that the quiescent absolute magnitude shows a marginal (but statistically significant) dependence on the duty cycle, with larger duty cycles producing a brighter quiescent state.

\section{Conclusion}\label{sec:conclusion}

The Outburst Catalogue provides apparent outburst and quiescent $V$ magnitudes, duty cycles, limits on the recurrence time, upper- and lower-limits on the distance and absolute quiescent magnitudes, colour information, orbital parameters, and X-ray counterparts where applicable for 722 dwarf novae (DN) and 309 other types of Cataclysmic Variable (CV), based on the Catalina Real-time Transient Survey (CRTS) $\sim9$ year light curves. These properties were determined by means of a classification script, presented in this paper. This is the largest sample of DN with estimates for these properties to date.

Using the Outburst Catalogue we have found correlations between the duty cycle and the orbital period, as well as the outburst recurrence time. The quiescent absolute magnitude shows a correlation with the orbital period, and with the duty cycle. We also show the range, and distribution, of the outburst properties and distances in the CRTS dwarf nova population. In a subsequent paper we will address the issue of completeness, and estimate the space density.

\section*{Acknowledgements}

Thank you to the referee, Michael Shara. The authors acknowledge funding from the Nederlandse Organisatie voor Wetenschappelijk Onderzoek and the Erasmus Mundus Programme SAPIENT. This paper uses data from the Catalina Sky Survey and Catalina Real Time Survey; the CSS is funded by the National Aeronautics and Space Administration under Grant No. NNG05GF22G issued through the Science Mission Directorate Near-Earth Objects Observations Program. The CRTS survey is supported by the U.S. National Science Foundation under grants AST-0909182.

This research has made use of NASA's Astrophysics Data System Bibliographic Services, as well as the SIMBAD database, operated at CDS, Strasbourg, France \citep{Wenger2000}.

Funding for SDSS-III has been provided by the Alfred P. Sloan Foundation, the Participating Institutions, the National Science Foundation, and the U.S. Department of Energy Office of Science. The SDSS-III web site is http://www.sdss3.org/. SDSS-III is managed by the Astrophysical Research Consortium for the Participating Institutions of the SDSS-III Collaboration including the University of Arizona, the Brazilian Participation Group, Brookhaven National Laboratory, University of Cambridge, Carnegie Mellon University, University of Florida, the French Participation Group, the German Participation Group, Harvard University, the Instituto de Astrofisica de Canarias, the Michigan State/Notre Dame/JINA Participation Group, Johns Hopkins University, Lawrence Berkeley National Laboratory, Max Planck Institute for Astrophysics, Max Planck Institute for Extraterrestrial Physics, New Mexico State University, New York University, Ohio State University, Pennsylvania State University, University of Portsmouth, Princeton University, the Spanish Participation Group, University of Tokyo, University of Utah, Vanderbilt University, University of Virginia, University of Washington, and Yale University.

This publication makes use of data products from the Wide-field Infrared Survey Explorer, which is a joint project of the University of California, Los Angeles, and the Jet Propulsion Laboratory/California Institute of Technology, funded by the National Aeronautics and Space Administration. We have made use of the ROSAT Data Archive of the Max-Planck-Institut für extraterrestrische Physik (MPE) at Garching, Germany. Data products from the Two Micron All Sky Survey, which is a joint project of the University of Massachusetts and the Infrared Processing and Analysis Center/California Institute of Technology, funded by the National Aeronautics and Space Administration and the National Science Foundation are used. We have also made use of data products obtained with XMM-Newton, an ESA science mission with instruments and contributions directly funded by ESA Member States and NASA. This work is based in part on data obtained as part of the UKIRT Infrared Deep Sky Survey. This research has also used data obtained from the Chandra Source Catalog, provided by the Chandra X-ray Center (CXC) as part of the Chandra Data Archive.

\bibliographystyle{mn2e.bst}
\bibliography{catalogue}

\label{lastpage}

\appendix
\section{Determining the distance upper-limit (Column 33)}\label{sec:distance2_uncertainties}

Empirically, the absolute magnitude of a DN in outburst ($V_{\rm O}$) is related to the orbital period ($P_{\rm orb}$) by
\begin{equation*}
 V_{\rm O} = 5.70 - 0.287 P_{\rm orb},
\end{equation*}
where $P_{\rm orb}$ is in hours \citep{Warner1987,Patterson2011}. This equation assumes a +0.8 mag correction to superoutbursts - larger amplitude outbursts that occur in a subclass of DN (the SU UMa stars) and have an additional source of emission possibly (believed to be from tidal heating, see \citet{Patterson2011} for a discussion). As \citeauthor{Patterson2011} points out, it is not clear whether this correction is appropriate.

As the sampling of the CRTS data is not sufficient to differentiate outbursts and superoutbursts, we use the form of this equation that does not assume a correction for superoutbursts \citep{Patterson2011},
\begin{equation}
 V_{\rm O} = 4.95 - 0.199 P_{\rm orb}\,.
 \label{eq:vmax-porb}
\end{equation}

The binary inclination ($i$) affects the observed $V_{\rm O}$, so to adjust for this, the correction for a flat, limb-darkened accretion disc from \citet{Paczynski1980} is applied:
\begin{equation}
 \Delta V_i=-2.5 \log((1+\frac{3}{2}\times\cos i)\times\cos i)\,.
 \label{eq:inclination_correction}
\end{equation}

Using the distance modulus, an estimate for the distance ($d$, in parsec) can thus be determined from $P_{\rm orb}$ and the apparent magnitude in outburst $v_{\rm O}$ via 
\begin{equation}
   d = 10^{(v_{\rm O}-V_{\rm O}+5)/5}\times 10^{-\Delta V_i/5}\,,
 \label{eq:distancemodulus}
\end{equation}
where the last term is the inclination-correction factor. As it is difficult to determine the inclination angle of a CV (e.g. \citealt{Littlefair2008}), few have $i$ estimates. Consequently we assume $i=56.7\degr$ (the average inclination) in cases where it is unknown.

We now discuss the uncertainties introduced by $i$, $v_{\rm O}$ and $V_{\rm O}$ to the distance estimate.

\subsection{Inclination}

\begin{figure}
  \centering
  \includegraphics[width=8cm]{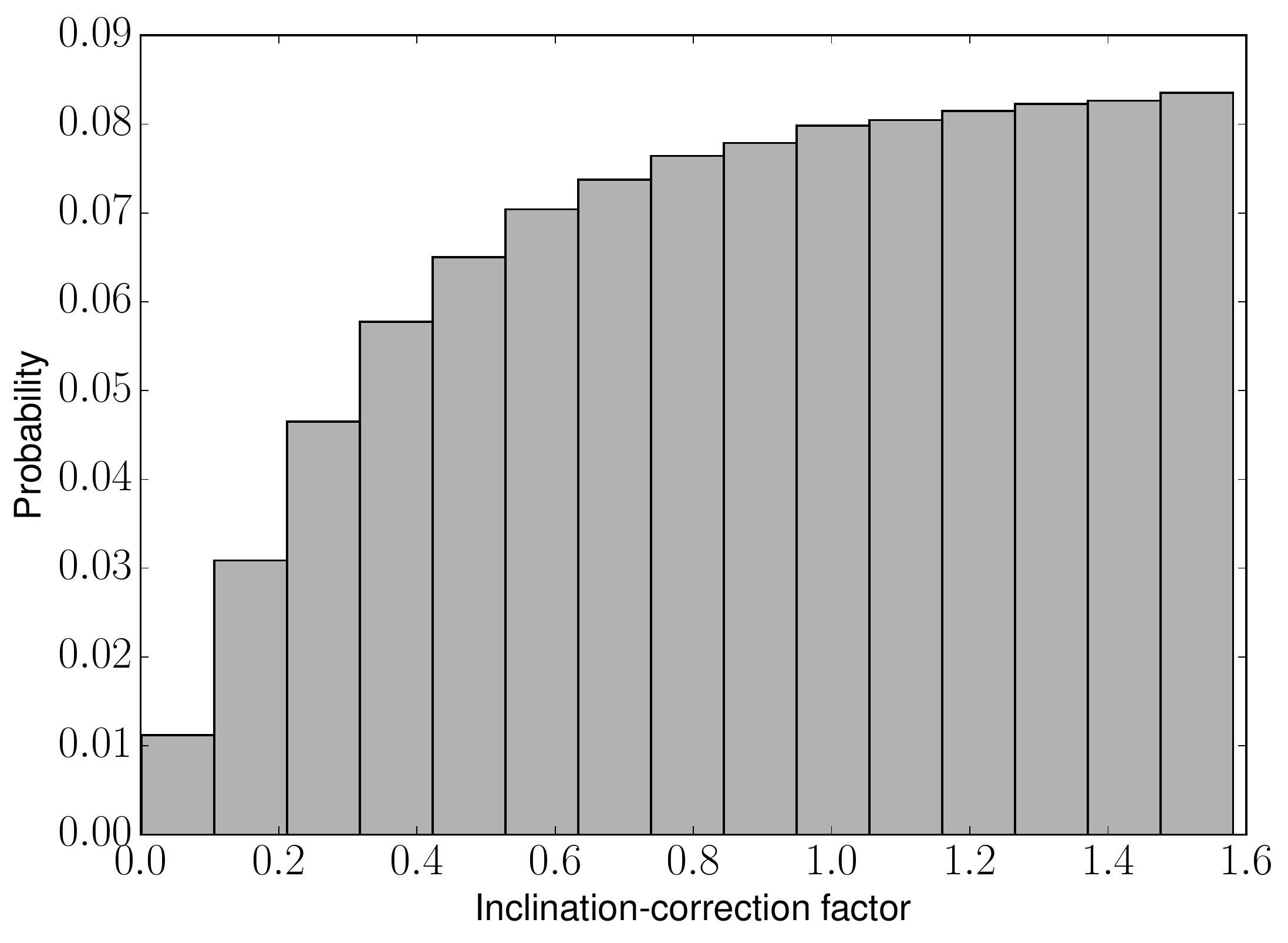}
  \caption{Probability that the estimated distance to a DN would need to be multiplied by a given inclination-correction factor to derive the true distance if an inclination of 56.7$\degr$ is assumed. The distance can be over- or under-estimated, but will always be less than a factor 1.58 times further, if only inclination effects are considered.}
  \label{fig:inclination_correction}
\end{figure}

Equation \ref{eq:inclination_correction} shows that the correction to $V_{\rm O}$ is highly sensitive on $i$. By assuming $i=56.7\degr$, we have set the inclination-correction factor equal to one. Taking a uniform distribution of $\cos i$, Figure \ref{fig:inclination_correction} shows that the true inclination-correction factor is in the range 0 to 1.58. Consequently the true distance is a factor 0 to 1.58 times the estimated distance.

The probability distribution is slightly skewed towards a closer distance than estimated, as there is a 55\% probability of an inclination-correction factor of less than 1. However, the probability decreases rapidly to smaller factors. Equation \ref{eq:inclination_correction} also becomes increasingly less reliable at large inclination angles (smaller inclination-correction factors), as the correction is for a flat disc and subsequently assumes an infinitely thin emission region at $i=90\degr$. At the high end, the distance can be underestimated by up to a factor 1.58 due to inclination effects.     

\subsection{Absolute Outburst Magnitude ($V_{\rm O}$)}

Equation \ref{eq:vmax-porb} was determined empirically and has a rms of 0.41 mag \citep{Patterson2011}. The uncertainty on $P_{\rm orb}$ is negligible, as it is typically known to an accuracy on the order of a few minutes or less. As mentioned previously, it is not clear if superoutbursts should have an additional correction to $V_{\rm O}$ and regardless, it is not possible to distinguish them from DN outbursts in the CRTS data. Equation \ref{eq:vmax-porb} does not assume a correction and hence is appropriate. 

\subsection{Quiescent Outburst Magnitude ($v_{\rm O}$)}

$v_{\rm O}^{\rm lim,f}$ is a lower-limit (faint-limit) for the outburst maximum $v_{\rm O}$, as the CRTS did not necessarily catch the DN at the peak of outburst. Outburst profiles and amplitudes vary between CVs and between outbursts of a single system, however the peak of outburst is typically a plateau phase that constitutes the most of the outburst. There is thus a large chance that it is detected at the peak, and this chance increases as more outbursts are sampled. As DN outbursts do not follow a standard template, it is not possible to estimate the uncertainty caused by assuming that $v_{\rm O}$ is the outburst maximum. However, since it is a faint-limit, the true distance will be closer than estimated.

Extinction will likewise make the estimate appear further than the true distance, as we cannot correct for it. As the CRTS does not observe within the galactic plane, it is not expected to produce a large uncertainty (in comparison to the inclination uncertainty).

\subsection{Upper-limit determination}

The distance estimate derived in this manner should only be used as a rough estimate. Based on these arguments, however, it is possible to make a more robust estimate for the upper-limit. The uncertainty on $v_{\rm O}$ produces an under-estimate of the distance, and the inclination and uncertainty on $V_{\rm O}$ give upper-limits for the true distance. Substituting the maximum inclination-factor of 1.58, and $V_{\rm O}=V_{\rm O}+0.41$ (the rms of Equation \ref{eq:vmax-porb} is 0.41) into Equation \ref{eq:distancemodulus} indicates that the true distance can be up to a factor 2 larger than estimated. An estimate for the upper-limit is thus be obtained by multiplying the distance estimate by a factor 2.

\end{document}